\documentclass[10pt]{article}
\usepackage{graphicx,psfrag,epsf}
\usepackage{enumerate}
\usepackage{pdfsync}
\usepackage{graphicx}
\usepackage[margin = 1 in]{geometry}
\usepackage{amsmath,mathtools}               
\usepackage{amsfonts}              
\usepackage{amsthm}                
\usepackage{amssymb}
\usepackage{color}
\usepackage{multirow}
\usepackage{tabularx}
\usepackage{caption}
\usepackage{tcolorbox}
\usepackage{setspace}
\usepackage[toc,page]{appendix}
\usepackage{algorithm}
\usepackage[noend]{algpseudocode}
\usepackage{natbib}
\usepackage{float}
\usepackage{hyperref}
\hypersetup{
    colorlinks=true,
    linkcolor=blue,
    filecolor=magenta,      
    urlcolor=blue,
}
\newtheorem{theorem}{Theorem}[section]

\newtheorem{condition}[theorem]{Condition}
\newcounter{example}[section]
\newenvironment{example}[1][]{\refstepcounter{example}\par\medskip
   \noindent \textbf{Example~\theexample. #1} \rmfamily}{\medskip}

\usepackage[T1]{fontenc}
\usepackage[utf8]{inputenc}
\usepackage{authblk}
\usepackage{lineno}

\newcommand{\xx}{\mathbf{x}}

\newcommand{\RR}{\mathbb{R}}

\newcommand{\T}{\mathcal T}
\newcommand{\Tc}{\mathcal T^c}

\newcommand{\cov}{\mathrm{cov}}
\newcommand{\var}{\mathrm{var}}

\newcommand{\E}{\mathrm E}

\newcommand{\Xx}{\mathbf{x}}

\newcommand{\sett}{\mathcal{T}}

\newcommand{\X}{\mathbf X}
\newcommand{\U}{\mathbf{U}}
\newcommand{\Ui}{\mathbf{U}_{(i)}}
\newcommand{\D}{\boldsymbol{\Lambda}}
\newcommand{\Vv}{\mathbf{V}}

\newcommand{\dn}{\triangleq}

\newcommand{\V}{\mbox{var}}

\newcommand{\Uh}{\bar{\mathbf U}_\ell}

\newcommand{\uu}{\mathbf u}
\newcommand{\pV}{\mathbb V}
\newcommand{\pLd}{\Upsilon}

\newcommand{\vv}{\mathbf v}

\newcommand{\half}{\frac{1}{2}}
\providecommand{\keywords}[1]{\textbf{Keywords:} #1}

\title{A Model-free Variable Screening Method \\ Based on Leverage Score}
 \author[1*]{Wenxuan Zhong}
 \author[2]{Yiwen Liu}
 \author[3]{Peng Zeng}
 \affil[1]{ Department of Statistics, University of Georgia, Athens, GA, 30602.}
 \affil[2]{ Department of Epidemiology and Biostatistics, University of Arizona, Tucson, AZ, 85724.}
 \affil[3]{ Department of Mathematics and Statistics, Auburn University, Auburn, AL, 36849.}
 \affil[*]{Corresponding author}
\date{}

\begin{document}
\maketitle
\doublespacing
\begin{abstract}
With rapid advances in information technology, massive datasets are collected in all fields of science, such as biology, chemistry, and social science. Useful or meaningful information is extracted from these data often through statistical learning or model fitting. In massive datasets, both sample size and number of predictors can be large, in which case conventional methods face computational challenges.
Recently, an innovative and effective sampling scheme based on leverage scores via singular value decompositions has been proposed to select rows of a design matrix as a surrogate of the full data in linear regression. Analogously, variable screening can be viewed as selecting rows of the design matrix. However, effective variable selection along this line of thinking remains elusive.
In this article, we bridge this gap to propose a weighted leverage variable screening method by utilizing both the left and right singular vectors of the design matrix. We show theoretically and empirically that the predictors selected using our method can consistently include true predictors not only for linear models but also for complicated general index models. Extensive simulation studies show that the weighted leverage screening method is highly computationally efficient and effective. We also demonstrate its success in identifying carcinoma related genes using spatial transcriptome data. 

\end{abstract}

\newpage
\section{Introduction}\label{sec:intro}
Among all 
statistical learning tools, regression analysis is one of the most popular methods and is widely used for modeling the relationship between a response $y$ and a series of predictors $x_1, \cdots, x_p$. Various models and methods have been developed for regression analysis in the literature, ranging from classic linear regression to nonparametric regression. Nevertheless, most regression models and methods can be seriously compromised if the dimensionality $p$ is large. It is ideal 
to select a subset of predictors to ensure the success of regression analysis.

    A wide range of variable selection methods have been proposed to facilitate dimension reduction in the literature, which can be mainly classified into two approaches: the subset selection approach including stepwise regression \citep{efroymson1960multiple}, forward selection, backward selection, etc; and the penalized likelihood approach including LASSO \citep{tibshirani1996regression}, non-negative garrotte \citep{breiman1995better,yuan2007non}, SCAD \citep{fan2001variable}, 
elastic net \citep{zou2005regularization}, 
penalized one-step estimator \citep{zou2008one}, and etc. Both of the two approaches can effectively regress $y$ on a selected subset of $x=(x_1,\ldots,x_p)^T$ when $x$ is of a moderate dimensionality. However, the aforementioned methods fail when $p$ is larger than the sample size $n$ \citep{fan2010selective}. 

For $p>>n$, an initial and deterministic screening step, which decreases the number of predictors from $p$ to $q$ where $q<<p$, can greatly improve computational efficiency. In many applications, we expect $q$ to be a rather crude upper bound to the number of ``true'' or ``predictive'' variables. Follow this line of thinking, a two-step screening strategy for linear regression was proposed by \cite{fan2008sure} to first screen out surely redundant variables and then refine the model using sophisticated variable selection methods.  In \cite{fan2008sure} and \cite{fan2009ultrahigh}, they developed a marginal correlation ranking method and showed $P(\T\subseteq A_q)\to 1$ under some conditions, where $\T$ is the subset of true variables and $A_q$ is selected subset of $q$ variables. The asymptotic performance of the screening methods was further studied in \cite{huang2008asymptotic}, \cite{hall2009tilting}, and \cite{hall2012using} under various settings. Despite the large number of available theoretical results, correlation ranking methods are only feasible when different variables are nearly independent. When the independence is not the case, the question that arises is how to screen predictors with moderate dependency structures. A simple solution has been proposed in \cite{wang2009forward} for linear models, showing that the forward selection procedure has screening consistency even when $p$ is substantially larger than $n$. However, the drawback of the forward selection method is its high computational cost. In addition, the aforementioned methods become ineffective when the underlying model is beyond linear. To address these issues, \cite{zhu2011model} extended the feature screening framework to semiparametric models. Their proposed procedure was demonstrated to possess ranking consistency, which leads to consistency in variable screening. \cite{li2012feature} developed a screening procedure based on distance correlation. Both methods consider the marginal relationship between each predictor and the response variable. \cite{zhou2019model} proposed cumulative divergence to characterize the functional dependence between predictors and the response variable, taking into account the joint effects among covariates during the screening process. These model-free methods are more robust but are often computationally intensive.

Heuristically, the screening process can be cast as a selection of columns of a data matrix. If we can find some ``importance score'' to evaluate a column's significance, we can screen out the insignificant columns with a probability that is calculated based on the importance score \citep{gallant1993nonlinear, mahoney2009cur}. This technique has been used extensively by computer scientists in finding a sparse matrix surrogate for a given matrix \citep{mahoney2008tensor, drineas2008relative, mahoney2009cur}. A leverage sampling method, in which rows and columns are sampled based on the leverage scores of data matrix $\X \in \mathbb R^{n\times p}$ and $\X^T$, has demonstrated much promise \citep{mahoney2009cur, ma2014statistical, ma2015leveraging} and is becoming the new research theme for matrix approximation. This method has recently been applied to linear regression problem to select a subsample,  i.e., select a set of rows of a data matrix.  Given $(\Xx^T_i, y_i)_{i=1}^{n}$ where $\Xx_i\in\mathbb R^{p}$, the linear regression model is of the form
\begin{equation}\label{eq:linear}
y_i=\Xx_i^T\beta+\epsilon_i,
\end{equation}
where $\beta\in \mathbb{R}^p$ is the regression coefficient that needs to be estimated,  and $\epsilon_i$ is the stochastic error that quantifies the measurement error. Let $\X=(\Xx_1, \cdots, \Xx_n)^T$. Without loss of generality, we assume $\X$ is centralized throughout this paper and has a rank $d$ singular value decomposition, i.e., $\X \approx \U \D \Vv^T$, where $\U\in\mathbb{R}^{n\times d}$, $\Vv\in \mathbb{R}^{p \times d}$ are column orthonormal matrices and $\D\in \mathbb{R}^{d\times d}$ is a diagonal matrix. Then, the importance of the $i$th observation or the $i$th row of $\X$ in a linear regression model is evaluated by its leverage score that is defined by $\U_{(i)}\U^T_{(i)}$ (or $||\U_{(i)}||^2_2$, where $||\cdot||_2$ denotes the $L_2$ norm), where $\U_{(i)}$ denotes the $i$th row of $\U$. Leverage scores are extensively used to measure how influential or important the rows of $\X$ are in a linear regression model. Recently, $\{ ||\U_{(i)}||^2_2 ,i=1,\ldots,n\}$ were used to select rows or subsample of $\X$ in a regression analysis such that the regression line obtained by the subsample can nicely approximate the regression line obtained by the full data \citep{ma2014statistical, ma2015leveraging}. In other words, the rows with large leverage scores are the rows that can be used to nicely approximate the regression line. 

Now returning to the variable screening problem, recall that selecting the columns of $\X$ can be cast as selecting the rows of $\X^T$.  Moreover, the leverage score of the $j$th row of $\X^T$ is defined by $||\Vv_{(j)}||^2_2$, where $\Vv_{(j)}$ denotes the $j$th row of $\Vv$. It can be considered as the influence of the $j$th column of $\X$ on the regression analysis. We thus intuitively use $||\Vv_{(j)}||^2_2$ as an ``importance score'' to sample the columns of $\X$ or the predictors. From this point on, we refer to $||\U_{(i)}||^2_2$ as the left leverage score and $||\Vv_{(j)}||^2_2$ as the right leverage score.  Analogous to left leverage score for selecting data points, right leverage scores might be used to select variables \citep{stewart1998four,drineas2006fast,dasgupta2007feature} when the regression model is linear. However, their performances are not as good as one may expect from this line of heuristic reasoning. The primary reason for the success of using the left leverage score for selecting the rows of $\X$ is that there is a theoretical  link between the left leverage score and response, i.e, 
\begin{equation*}
    \frac{\partial \hat{y}_i}{\partial y_i}= ||\U_{(i)}||^2_2,
\end{equation*}
where $\hat{y}_i$ is the $i$th fitted value of least squares.  That is, the left leverage score measures the changes of the fitted value of the response with respect to a small change of the response.  It remains elusive whether there exists some theoretical underpinning for linking the right leverage score and response.
More importantly, in practice, the relationship between the response variable and predictors is usually more complicated than a linear model, which adds another layer of complications in developing the leverage screening approach. It is conceivable that the development of variable screening or variable selection methods based on the right leverage score  when the underlying models are beyond linear models is very challenging.
Their theoretical underpinning remains unknown even for fixed $p$ if there is no concrete model to associate response and predictors, because there is no unified likelihood function to study their statistical properties. The problem may be even harder for growing $p$ or even $p>>n$. 

To surmount these challenges, in this article, we propose a variable screening criterion that is derived by integrating both the right leverage score $||\Vv_{(j)}||^2_2$ and left leverage score $||\U_{(i)}||^2_2$  together to evaluate columns' or predictors' importance in regression analysis. More specifically, we assume that given  $k$ linear combinations of predictors $x$, response variable $y$ and predictors $x$ are independent. Our method is ``model-free" in the sense that there is no explicit link function between  $y$ and $x$. 
We develop a weighted leverage score to measure the ``importance'' of each variable in the model.
Based on the score, we design a one-pass variable screening algorithm. 
More importantly, we develop a BIC-type criterion to decide the number of selected predictors.  We show empirically and theoretically that our proposed method can consistently select the non-redundant predictors. 

Our main methodological contribution is to develop a variable screening method in high dimensional  model-free setting.  Compared with the variable screening methods for parametric models, our method avoids the model mis-specification error. Compared with the variable selection in more flexible nonparametric models \citep{ravikumar2009sparse,fan2011nonparametric}, our method does not estimate the unknown link function between response and predictors and has substantial analytical and computational advantages.  The proposed weighted leverage score is calculated using the singular value decomposition, which can be found in most computing software. To the best of our knowledge, our work is the first to relate the leverage score with variable selection in semi-parametric models.
The screening algorithm is a one-pass algorithm, which is scalable to high dimensional settings. We also develop BIC-type criteria to select the number of variables.
Our main theoretical contribution is to establish screening consistency under very general regularity conditions. In particular, we show that the weighted leverage scores of the true predictors are larger than those of the redundant predictors. Moreover, the BIC-type criteria we develop are consistent for variable screening.

The rest of the paper is organized as follows. In Section \ref{sec:wls}, we briefly review the general multiple index model and introduce the motivation of using weighted leverage score (WLS) for variable screening. Section \ref{sec:thy} illustrates the asymptotic behavior and rank consistency of WLS. Several implementation issues of the procedure are discussed in Section \ref{sec:imp}. Simulation studies and a real data example are reported in Section \ref{sec:sim} and Section \ref{sec:app}. Section \ref{sec:dis} concludes the paper with a discussion. All proofs are provided in Supplementary Material.

\def \Ss {\mathcal{S}}
\def \sSs {\Ss}
\section{Model-free variable screening using weighted leverage score}\label{sec:wls}
\subsection{Model-free regression and sufficient dimension reduction }\label{sir}
 
Without loss of generality, we assume from this point on that $x$ is a $p$-dimensional random vector with mean zero and variance-covariance $\Sigma$, and $y \in \mathbb R$ is the response variable. 
Let $\Ss$ be a subspace of $\RR^p$, and $P_\Ss$ be the projection operator from $\RR^p$ to $\Ss$ in 
an inner product space. If
\begin{equation}
y\perp x|P_\Ss x,
\label{eq1}
\end{equation}
where $\perp$ means ``independent of'', it is said that $P_\Ss x$ is sufficient for the dependence of $y$ on $x$ \citep{cook1995introduction, cook1996graphics, cook1998principal}. In other words, the projection $P_\Ss x$ captures all the information contained in $x$ regarding $y$. Regressing $y$ on $x$ thus is equivalent to regressing $y$ on $P_\Ss x$. A dimension reduction is achieved if the dimensionality of $\Ss$ is smaller than $p$.  

Expression (\ref{eq1}) does not spell out any model, i.e., ``model-free'', in the sense of classical regression settings, where the conditional distribution of $y$ given $x$ is explicitly stated. 
However, it is equivalent to a general index model proposed in \cite{li1991sliced},
\begin{equation}
y = f(\beta^T_1 x, \ldots, \beta^T_k x, \epsilon),
\label{eq2}
\end{equation}
where $f(\cdot)$ is an unknown function, $\beta_1,\ldots,\beta_k$ are $p$-dimensional vectors, $k$ is an integer much smaller than $p$, and $\epsilon$ is a stochastic error. It is easy to show that $y$ and $x$ in model (\ref{eq2}) are independent if $\{\beta^T_m x|m=1, \cdots, k\}$ are given. Therefore, the subspace spanned by $\{\beta^T_m x|m=1, \cdots, k\}$ can serve as the subspace in model (\ref{eq1}). Conversely, if (\ref{eq1}) holds, there exist $f(\cdot)$ and $\epsilon$ such that (\ref{eq2}) holds. A brief proof of the equivalence between the two models can be found in \cite{zeng2010integral}.

Model (\ref{eq1}) and (\ref{eq2}) are referred to as the sufficient dimension reduction (SDR) regression model, and $\Ss$ is referred to as a dimension reduction subspace. Dimension reduction subspace may not be unique. \cite{cook1996graphics} introduced an important concept called {\it central subspace}, which is defined as the intersection of all dimension reduction subspaces when it is a dimension reduction subspace itself. The central subspace is denoted by $\Ss_{y|x}$, and the dimension of $\Ss_{y|x}$ is called the structural dimension of regressing $y$ on $x$. Under mild conditions, it can be shown that $\Ss_{y|x}$ exists (see \cite{cook1995introduction} for details). Throughout this paper, we assume the existence of $\Ss_{y|x}$.

The dimension reduction regression model is unarguably the most general formulation and covers a wide range of parametric and semi-parametric models. For example, if $y$ is a discrete variable taking values in $\{ 1$, $2$, $\cdots$, $K\}$, the dimension reduction regression model covers logistic regression and many classification models. If $y$ is a continuous variable taking values in $\RR$,  linear regression model, partial linear model, and single index model are its special cases. Comparing to existing models, the dimension reduction regression models not only provide a much flexible model structure to address the nonlinear dependency but also keep the model simplicity. Thus it has been extensively used to analyze the complicated high-dimensional data. Despite the popularity of the SDR in high-dimensional regression, it has been shown in \cite{zhu2006sliced} that the efficiency of the estimates in the SDR model deteriorates when one includes more and more irrelevant features (covariates). Thus, instead of identifying the low dimensional projections, simultaneously detecting the non-redundant predictors is more critical especially when $p>>n$.

\def \nh {\frac{1}{n_h}}
\def \by {\mathbf{y}}
\def \halfn {\frac{n}{2}}
\newcommand{\y}{\boldsymbol y}
\def \crr {\mbox{corr}}
\def \tm {\tilde{\mathcal{M}}}

\subsection{Weighted leverage score for model-free regression}

\def \tyi {\tilde{Y}^{(i)}}
\def \tz {\tilde{Z}}
\def \tgammai {\tilde{\gamma}^{(i)}}
\def \tepsiloni {\tilde{\epsilon}^{(i)}}

Given $(\xx_i^T, y_i)$ for ${i=1}, \ldots, {n}$, notice that $\xx_i$ can be approximated by $\Vv \D \Ui^T$. Recall that $\Ui$ denotes the $i$th row of left singular matrix $\U$, and it has a natural connection with the response variable $y_i$ as it contains the sample information of the data. To reflect such connection when constructing the weighted leverage score, we integrate both $\Ui$ and $y_i$ together by utilizing the slicing scheme and inverse regression idea. We first divide the range of the response variable into $h$ intervals or slices $S_1, \cdots, S_h$ and then group the $\Ui$ accordingly if its corresponding $y_i$ falls into the same slice. For each slice, we calculate its slice mean by taking its group mean $\Uh = \frac{1}{n_\ell}\sum_{i=1}^{n}\Ui I(y_i\in S_\ell)$, where $I(\cdot)$ is the indicator function, and $n_\ell=\sum_{i=1}^{n} I( y_i \in S_\ell)$ for $\ell=1,\ldots,h$. Finally, we calculate the sample variance of the slice means to obtain an estimate of $\V[\E(\Ui|y_i)]$ as $\sum_{\ell=1}^{h} \frac{n_\ell}{n} \Uh^T\Uh$. 
The matrix $\V[\E(\Ui|y_i)]$ captures the information contained in the link function $f$ of model (\ref{eq2}).
Further, $\Vv_{(j)}$, as the $j$th row of the right singular matrix $\Vv$, reflects the predictor information. Thus, to evaluate how influential a predictor is to the regression model (\ref{eq2}), we propose the weighted leverage score of $j$th predictor $\hat\omega_j$ as the right leverage score $\|\Vv_{(j)}\|_2$ weighted by a matrix formulated based on the left singular matrix $\U$, 
\begin{equation}\label{wls}
\hat \omega_j \dn \Vv_{(j)}(\sum_{\ell=1}^{h} \frac{n_\ell}{n} \bar\U_\ell^T\bar\U_\ell)\Vv_{(j)}^T.
\end{equation}
The weighted leverage score is constructed on the slicing scheme and is closely related to the slice inverse regression (SIR) method proposed in \cite{li1991sliced}. 
It has been shown in \cite{li1991sliced} that when the linearity condition is satisfied, the inverse regression curve $\E(\xx_i|y_i)$ resides in the space that is spanned by $\beta_1\Sigma, \cdots, \beta_k\Sigma$. Thus $P_\Ss=(\beta_1\Sigma, \cdots, \beta_k\Sigma)$ is the basis of the space that contains $\E(\xx_i|y_i)$. Based on this fact, \cite{li1991sliced} proposed to estimate $\beta_1,\ldots,\beta_k$ by conducting eigenvalue decomposition on $\V[\E(\Sigma^{-\half} \xx_i|y_i)]$.
%
%
%
Now the key to the success of dimension reduction is how to estimate $\V[\E(\Sigma^{-\half} \xx_i|y_i)]$. Notice that the inverse regression curve $\E(\xx_i|y_i)$ is a function of a one dimensional response variable $y_i$, it thus can be easily approximated by a step function. 
More specifically, we can estimate $\E(\Sigma^{-\half}\xx_i|y_i)$ by $n^{(-1)}_\ell\sum_{i=1}^{n} \hat\Sigma^{-\half}\xx_i I(y_i\in S_\ell)$, where $\hat\Sigma$ is an estimator of $\Sigma$.
%
Further, with $\hat\Sigma = \Vv\D^2\Vv^T$, we can write $\hat\Sigma^{-\half} \xx_i$ as $\Vv \Ui^T$. Then $\V[\E(\Sigma^{-\half} \xx_i|y_i)]$ is estimated by
\begin{equation}
 \Vv (\sum_{\ell=1}^{h} \frac{n_\ell}{n} \bar\U_\ell^T\bar\U_\ell) \Vv^T, 
 \label{dhat}
\end{equation}
of which the diagonal elements are the weighted leverage scores. In the next Section, we show that the weighted leverage scores can consistently select the true predictors for fixed $S_1, \cdots, S_h$.
%
%

Intuitively, the $\hat\omega_j$ can be cast as a weighted right leverage score ($||\Vv_{(j)}||_2$), where the weights are constructed by the left singular matrix $\U$. We thus refer to $\hat \omega_j$ as the weighted leverage score. Notice that the weight matrix, formulated by $\U$ and $\{y_i\}_{i=1}^n$, captures the nonparametric information $f$. It is the same for all predictors when constructing $\hat\omega_j$'s. While $\Vv_{(j)}$ captures the predictor-specific information. Thus the weighted leverage score can be naturally used to evaluate a predictor's significance in model (\ref{eq2}). 
Using the weighted leverage score, we propose a simple variable screening algorithm that is sketched in Algorithm 1. 
\begin{algorithm}[H]
  \caption{The weighted leverage score screening algorithm}
  {\bf Step 1}. For $j = 1, \ldots, p$, calculate the weighted leverage score of $j$th variable, $\hat \omega_j$, by equation (\ref{wls}). 
  
  {\bf Step 2}. Sort the weighted leverage scores in decreasing order and denote them as $\hat \omega_{(1)} > \ldots > \hat \omega_{(p)}$. Output the predictors that with the highest $p_0$ weighted leverage scores. The final selected predictor set is
  \vspace{-20pt}
  \begin{eqnarray*}
  \mathcal A = \{ j: \ \ \hat \omega_j \ge \hat \omega_{(p_0)} \}.
  \end{eqnarray*} 
  \vspace{-20pt}
\end{algorithm}
  
\section{Theoretical Justification}\label{sec:thy}
In this section, we show that the minimum weighted leverage score of true predictors is larger than the maximum weighted leverage score of redundant predictors. Consequently, the true predictors are first selected if we rank the predictors according to their weighted leverage scores. We demonstrate that this ranking property holds for both the population and sample weighted leverage scores.


Let us first consider the ranking property of the population weighted leverage score, denoted as $\omega_j$. Clearly, $\omega_j$ is the $j$th diagonal element of $\pV (\sum_{\ell=1}^{h} p_\ell \uu_\ell \uu_\ell^T )\pV^T,$
where $\pV$ and $\uu_\ell=E(\uu_i | y_i \in S_\ell)$ are the population version of $\Vv$ and $\bar{\U}_\ell$ respectively, and a rigorous definition of $\uu_i$ can be found in condition \ref{cond:representx}. For a fixed slicing scheme $\{S_\ell\}_{\ell=1}^{h}$, we have $p_\ell=P(y\in S_\ell)$. Under certain regularity conditions, we first show that the minimum $\{\omega_j|j\in \sett\}$ is larger than the maximum $\{\omega_j|j\in\sett^{c}\}$, where $\sett$ is the collection of $p_0$ true predictors under model (\ref{eq1}) and (\ref{eq2}), and $\{\cdot\}^c$ denotes the complement of a given set.

To ease the description, we introduce the following notations. Let $\lambda_{\max}(\cdot)$ and $\lambda_{\min}(\cdot)$ denote the functions that take the maximum and minimum eigenvalues/singular values of a matrix respectively. Let $V_h=\sum_{\ell=1}^{h}p_\ell\E(x| y \in S_\ell)\E( x | y\in S_\ell)^T$ and $M_{hk}=B^TV_hB$, where $B=(\beta_1, \cdots, \beta_k)$ in model (\ref{eq2}). Clearly $V_h$ is an estimate of $\V[\E(x| \tilde y)]$ and $M_{hk}$ is an estimate of $\V[\E(B^T x| \tilde y)]$ when $\E(x)=0$, where $\tilde y$ is discretized $y$. To prove the ranking property of $\omega_j$, we require the following conditions on the random vectors $x$, from which the left and right singular vectors are derived.

\begin{condition}\label{cond:linear}
	Assume that $x$ is from a non-degenerate elliptically symmetric distribution.
\end{condition}

\begin{condition}\label{cond:Sigma}
	There exist two positive constants $\tau_{\min}$ and $\tau_{\max}$, such that $ \tau_{\min} \le \lambda_{\min}(\Sigma) \le \lambda_{\max}(\Sigma) \le \tau_{\max}$.
\end{condition}

\begin{condition}\label{cond:evV}
 For fixed slicing scheme, $S_1, \cdots, S_h$, there exist two positive constants $\tau^h_{\min}$ and $\tau^h_{\max}$ such that $\lambda_{\max}(V_h) \le \tau^h_{\max}$ and $\lambda_{\min}(M_{hk}) \ge \tau^h_{\min}$.
\end{condition}

\begin{condition} \label{cond:beta}
There exists a positive constant $\mathcal C_0 > 0$ such that for $j \in \T$,
\begin{eqnarray*} \label{ineq:SN-ratio}
 \min_{j \in \sett} \| B_{(j)} \|_2
 > \mathcal C_0 \frac{ \lambda_{\max}[ \cov(x_{\T^c},x_{\T}) ] }{\lambda_{\min}[\cov(x_{\T},x_{\T})] },
 \end{eqnarray*}
where $B_{(j)}$ is the $j$th row of the $p\times k$ coefficient matrix $B$.
\end{condition}

Condition \ref{cond:linear} is also called the design condition and was first proposed in \cite{duan1991slicing} to ensure that $\beta_1, \cdots, \beta_k$ in model (\ref{eq2}) are the eigenvectors of $\V[\E(x|y)]$. It is slightly stronger than the linearity condition that was required in \cite{li1991sliced}. If condition \ref{cond:linear} holds, we have $E(x | B^T x ) = \cov( x, B^T x) B^T x$. The conditional expectation of $x$ given $B^Tx$ is linear in $B^Tx$. The design condition was also required in \cite{zhu2011model} to establish variable screening consistency. It always holds if $x$ follows a multivariate Gaussian distribution, a condition that is required by most variable selection procedures. Condition \ref{cond:Sigma} is imposed on the population covariance matrix, which ensures that no predictor has a dominate variance or is linearly dependent on other predictors \citep{zhong2012correlation}. Condition \ref{cond:evV} ensures that no $\E(x_{j_1}|y)$ or $\E(\beta_{m_1}^T x|y)$ has a dominate variance or is linearly dependent on $\E(x_{j_2}|y)$ or $\E(\beta_{m_2}^T x |y)$ respectively for $j_1\neq j_2$ and $m_1 \neq m_2$. This condition is slightly stronger than the so-called coverage condition \citep{cook2004testing} that ensures $V_h$ to recover all the SDR directions. Condition \ref{cond:Sigma} and \ref{cond:evV} are necessary conditions. Without the two conditions, neither $\Ss$ is well defined nor $V_h$ can be used to recover model (\ref{eq1}) and (\ref{eq2}). Similar conditions were also required in \cite{li1991sliced} and \cite{zhong2012correlation} to ensure the consistency of $B$. Condition~\ref{cond:beta} is a sufficient condition for the success of Theorem~\ref{thm:truemodel} (Supplementary Material S.3.1). It requires that the coefficients of true predictors are large enough to be detectable. Intuitively, the projection of the redundant variables on the space that spanned by the true predictors must be smaller than the projection of the response $y$ on the space that is spanned by the true predictors. It is easy to see that condition~\ref{cond:beta} always holds when $x_{\T}$ and $x_{\T^c}$ are independent.
\begin{theorem}\label{thm:truemodel}
Given conditions \ref{cond:linear}-\ref{cond:beta} are satisfied, we have $\min_{j \in \sett} \omega_j > \max_{j \in \sett^c} \omega_j$.
\end{theorem}
Theorem \ref{thm:truemodel} implies that the weighted leverage score of any true predictor is larger than that of any redundant predictors. The proof of this theorem is collected in Supplementary Material S.1.1. If $\max_j\mid\hat\omega_j-\omega_j\mid$ is smaller than $\delta=\min_{j \in \sett} \omega_j-\max_{j \in \sett^c} \omega_j$, we thus have that $\hat\omega_j$ possesses the ranking consistency.

If we further assume that the following conditions are satisfied, we showed that $\hat \omega_j$ still has the ranking property when $p>>n$ as both $n$ and $p$ go to infinity.

\begin{condition}\label{cond:representx}
Assume $\xx_1,\ldots,\xx_n$ are i.i.d. p-dimensional random vectors with the representation
\begin{equation}\label{eq:xrepresent}
\xx_i =  \pV \pLd \uu_i,
\end{equation}
where $\pV = ( \vv_1, \ldots, \vv_p) \in \mathbb R^{p\times p}$ with $\vv_j$ being the $j$th eigenvector of $\Sigma$, $\pLd=\text{diag }(\lambda_1, \ldots, \lambda_p) \in \RR^{p \times p}$ with $\lambda_j$ being the square root of $j$th eigenvalue of $\Sigma$, and $\uu_i=(u_{i1}, \ldots, u_{ip})^T$ with each element be i.i.d. sub-Gaussian random variable with zero mean and unit variance.
\end{condition}

\begin{condition}\label{cond:multicomp}
  Assume the spiked model such that $\lambda_1 > \ldots > \lambda_d >> \lambda_{d+1} \ge \ldots \ge \lambda_p > 0$. The spiked eigenvalues are well separated and $\lambda_{j}^2/\lambda_{i}^2=c_{ji}$ for $ i,j \in \{1,\ldots,d\}$ and $i\neq j$, where $c_{ji}$ is a positive constant. The non-spiked eigenvalues are bounded by some positive constants.
\end{condition}

\begin{condition}\label{cond:multicomp2}
  Assume $p>n$. For spiked eigenvalues $\{\lambda^2_j\}_{j=1}^d$, $p/(\sqrt{n}\lambda_j^2) \rightarrow 0$. For non-spiked eigenvalues $\{\lambda^2_j\}_{j=d+1}^p$, there exists a positive constant $\bar c$ such that  $(p-d)^{-1} \sum_{j=d+1}^p \lambda^2_j = \bar c + o(n^{-1/2})$. 
\end{condition}

\begin{condition} \label{cond:ui}
Given any slice $\{S_\ell\}_{\ell=1}^h$, $\E( u_{ij} |y_i\in S_\ell) =0$ for $j=d+1,\ldots,p$, and $\E( |u_{ij}|^4 |y_i\in S_\ell) < \infty$ for $j=1,\ldots, d$.
\end{condition}

In condition \ref{cond:representx}, we assume $u_{i1}, \ldots, u_{ip}$ are i.i.d. sub-Gaussian random variables. Given the variance-covariance matrix $\Sigma$, then $\xx_i$ having the representation is also sub-Gaussian distributed with strong tail decay. Compared with condition 3.1 that requires a symmetric distribution, this condition emphasizes on the tail behavior of the distribution of $\mathbf{x}_i$. This class of distributions is sufficiently wide enough to contain all bounded distributions.

Condition \ref{cond:multicomp} assumes the spike covariance model introduced by \cite{johnstone2001distribution}. The eigenvalues of covariance matrix are divided into distinguishable spiked ones and bounded non-spiked ones. A similar condition can be found in \cite{shen2014general,shen2016statistics} and \cite{fan2015asymptotics}. The well separated spiked eigenvalues satisfy $\min_{j\le d}(\lambda^2_j - \lambda^2_{j-1})/\lambda^2_j \ge c_0$ for some $c_0>0$. The non-spiked ones are bounded by two positive constants $c_l$ and $c_u$ such that $c_l\le \lambda^2_j \le c_u$ for $j>d$.

The spiked covariance model typically assumes that several large eigenvalues are well-separated from the remaining. In this paper, we are particularly interested in the spiked part since the corresponding directions explain most of the variations in the data, while the remaining directions contain noise. Since the weighted leverage score is developed based on both the left and right singular matrices, to control the signal and noise contained in the data, we assume in condition 3.7 that the first $d$ directions explain a large proportion of the information of the data, represented by $(\sum_{j=1}^d \lambda^2_j) / (\sum_{j=1}^p \lambda^2_j)$. Here we consider $d$ as a fixed number and is independent of $n$ and $p$, which means that $d << n$ as $n \rightarrow \infty$. Furthermore, $d$ is also independent of the number of true predictors $p_0$.

Condition~\ref{cond:multicomp2} allows $p/n \rightarrow \infty$ in a way such that $\{\lambda_j\}_{j=1}^d$ also grows fast enough to ensure $p/(\sqrt{n}\lambda_j^2)$ goes to zero. The same condition was required in \cite{fan2015asymptotics} to guarantee a clear separation of the signal from the noise. 
Together with conditions 3.7 and 3.8, we may establish the asymptotic behaviors of the spiked eigenvalues and corresponding eigenvectors. 
An example of such spiked model could have eigenvalues $\lambda_1^2 > \ldots > \lambda_d^2 > 1 = \ldots = 1$,  where $\lambda_1^2, \ldots, \lambda_d^2$ are spiked eigenvalues, and the rest are non-spiked eigenvalues.
Condition~\ref{cond:ui} requires that the conditional expectation $\E(\xx_i|y_i\in S_\ell)$ is contained in the space spanned by $\vv_1,\ldots,\vv_d$ with $\lambda_1\E(u_{i1}|y_i\in S_\ell), \ldots,\lambda_d \E(u_{id}|y_i\in S_\ell)$ as coefficients.

\begin{theorem}\label{thm:consist}
Assume conditions \ref{cond:linear}-\ref{cond:beta} and \ref{cond:representx}-\ref{cond:ui} are satisfied. Denote $\delta = \min_{j \in \mathcal T} \omega_j - \max_{j \in \mathcal T^c} \omega_j $. There exists a positive constant $\mathcal C_0$ and $\xi$ such that for $\xi \in ( \mathcal C_0 \frac{p}{\sqrt{n} \lambda_d^2}, \delta/2 )$,
\begin{equation}
P( \max_{1\le j\le p} |\hat\omega_j - \omega_j| < \xi ) \rightarrow 1.
\end{equation} 
In addition,
\begin{equation}
	P( \min_{j\in \T} \hat \omega_j > \max_{j \in \Tc} \hat \omega_j ) \rightarrow 1.
\end{equation}
\end{theorem}
%
The proof of Theorem \ref{thm:consist} is collected in Supplementary Material S.1.2.

\section{Implementation Issues} \label{sec:imp}
There are two challenges in implementing the WLS algorithm: 1) The specification of the number of spiked eigenvalues $d$ is crucial for detecting the amount of signals; 2) The specification of the number of selected predictors significantly affects the false selection and false rejection and consequently is another critical issue in practice. 
In the following, we discuss how to deal with these two issues. 
\subsection{Decide the number of spiked eigenvalues $d$}
By analyzing the eigenvalues of the covariance matrix, we suggest a BIC-type of criterion for determining the number of spiked eigenvalues $d$. Let $\theta_i =\lambda_i^2/\lambda_1^2+1$ and $\hat\theta_i = \hat\lambda_i^2/\hat\lambda_1^2+1$, where $\lambda_i^2$ and $\hat\lambda_i^2$ are $i$th eigenvalues of $\Sigma$ and $\hat\Sigma$ respectively for $i=1,\ldots,\min(n,p)$. It is clear that $\hat\theta_1 > \ldots > \hat \theta_d > \ldots > \hat\theta_{\min(n,p)}$. Let $r$ be the number of current selected spiked eigenvalues, we define a criterion of BIC-type as follows.
\begin{equation}\label{eq:BIC}
D(r) = -\sum_{i=r+1}^{\min(n,p)} ( \log \hat\theta_i + 1 - \hat\theta_i ) + c_{n_1} r/n^{\half},
\end{equation}
where $c_{n_1}$ is a positive constant. The estimator of $d$ is defined as the minimizer $\hat d$ of $D(r)$ over $r=1,\ldots, \min(n,p)$.
Notice that the first term of (\ref{eq:BIC}) indicates the loss of information. It decreases as we include more eigenvalues. When $r>d$, the decrease in the loss of information becomes smaller than the penalty, and $D(r)$ starts to increase. 
The following theorem states the consistency of $\hat d$.
\begin{theorem}\label{thm:selectd}
Assume conditions \ref{cond:representx}-\ref{cond:ui} are satisfied. Let $\hat d = \arg_r \min D(r)$, we have $P(\hat d = d) \rightarrow 1$.
\end{theorem}
Theorem~\ref{thm:selectd} ensures that $D(r)$ is consistent for specifying $d$. The proof of Theorem~\ref{thm:selectd} is collected in Supplementary Material S.1.3. Our simulation study shows that the proposed criterion leads to the correct specification of $d$ and can be generally used in practice.
In terms of calculating singular values, we consider the reduced singular value decomposition (SVD) in the $p>n$ scenario in this paper. The $n$ largest singular values are calculated first, and the number of spiked eigenvalues $\hat{d}$ is then determined using this criterion. We calculate the weighted leverage scores based on the first $\hat{d}$ singular vectors. For ultra-high dimensional data, we recommend using fast algorithms for SVD, such as the randomized block Krylov method \citep{musco2015randomized},  the fast stochastic k-SVD algorithm \citep{shamir2016fast}, and the LazySVD \citep{allen2016lazysvd}.

\subsection{Decide the number of predictors}\label{sec:sub2}
Theorem \ref{thm:consist} ensures that the weighted leverage scores preserve the ranking consistency under certain conditions. To achieve the screening consistency, we rank each predictor's WLS and keep $p_0$ predictors with the largest WLS. A good estimate of $p_0$ thus is critical for screening consistency. When $\hat p_0$ is too large, we keep too many redundant predictors, and if $\hat p_0$ is too small, we miss a lot of true predictors. In literature, a common criterion to decide $\hat p_0$ is the BIC-type criterion that was used in \cite{chen2008extended} and \cite{wang2009forward}. In this article, we propose a modified version of BIC-type criterion. Under some conditions, we show that the subset of predictors that minimizes the modified BIC-type criterion consistently includes the true predictors. Next, we introduce the modified BIC-type criterion. 

Arrange the predictors such that $\hat \omega_{1} > \ldots > \hat \omega_{p}$ is satisfied. Let $r$ be the number of currently selected predictors. Similar as BIC, we define 
\begin{equation}\label{eq:bic}
 G(r) = - \log( \sum_{j=1}^{r} \hat \omega_{j}) + r(\log n + c_{n_2}\log p)/\max(n,p),
\end{equation}
where 
$c_{n_2}$ is a pre-specified positive constant. Notice that $G(r-1)-G(r)= \log(1 + \hat \omega_{r}/\sum_{j=1}^{r-1} \hat \omega_j)- (\log n + c_{n_2}\log p)/\max(n,p)$. The less significant the $r$th predictor is, the smaller the $\hat\omega_{r}$ is. The value of $G(r-1)-G(r)$  thus is smaller when adding the $r$th predictor, until to some point that $\hat\omega_{r}$ is too small to have positive $G(r-1)-G(r)$, $G(r)$ starts to increase.
We show in Theorem \ref{thm:bic} that $G(r)$ can consistently screen out the redundant predictors. 


\begin{theorem}\label{thm:bic}
Assume that conditions \ref{cond:linear} - \ref{cond:beta} hold. If we further assume that conditions \ref{cond:representx} - \ref{cond:ui} are satisfied, we have
\begin{equation}
P(\sett \subset \mathcal{A} ) \rightarrow 1,
\end{equation}
where $\sett$ is the subset of true predictors and $\mathcal{A}$ is the subset of selected predictors that minimizes $G(r)$.
\end{theorem}
\noindent The proof of Theorem \ref{thm:bic} is collected in Supplementary Material S.1.4. Theorem \ref{thm:bic} ensures that $G(r)$ is consistent for predictor screening. In Section \ref{sec:sim}, we use comprehensive simulation studies to justify the empirical performance of $G(\cdot)$ in determining the model size.

\section{Simulation Study}\label{sec:sim}
We have conducted extensive simulation studies to compare the performance of WLS screening method with that of existing variable screening methods, including sure independence ranking and screening (SIRS) \citep{zhu2011model} and sure independence screening with distance correlation (DC-SIS) \citep{li2012feature}. 
The performances of the aforementioned variable screening methods were evaluated by the following four criteria: the average number of irrelevant predictors falsely selected as true predictors (denoted by FP), the average number of true predictors falsely excluded (denoted by FN), the average minimum model size to include all true predictors (denoted by $\mathcal M$), and  CPU time charged for the execution of the corresponding method. We used $[n/\log(n)]$ as the cutoff for SIRS and DC-SIS, and $G(\cdot)$ to determine the number of selected predictors for WLS.  The FP and FN were used to examine the accuracy of variable screening procedures. The $\mathcal M$ is an indicator of the ranking property with a smaller value indicating a better screening process. The computation time was also recorded here for the evaluation of efficiency.


Throughout this section, we used the following two settings to generate i.i.d. copies of $x$. (1) Assume $x=(x_1,\ldots,x_p)^T$ and let the index set of the true predictors be $\mathcal I_{\T}=\{t_1=1, t_2=10, t_3=15, t_4=20, t_5=25, t_6=30\}$. We generated i.i.d. copies of $x$ by $\xx_i = \pV \pLd \uu_i$ for $i=1,\ldots,n$, where $\pV$ is a $p$-by-$p$ orthonormal matrix, $\pLd=\text{diag}(\lambda_1, \lambda_2,\ldots,\lambda_d,1,\ldots,1)$ has $d$ spiked values, and $\uu_i$ follows a multivariate normal distribution with $\E(\uu_i)=\mathbf 0$ and $\var(\uu_i)=I_p$. 
(2) We further studied the performance of WLS when the covariance matrix $\Sigma$ does not have spiked eigenvalues. Assume that $x=(x_1,\ldots,x_p)^T$ follows a multivariate normal distribution with mean zero and covariance $\mbox{Cov}(x_i,x_j) = \rho^{|i-j|}$ and let the index set of true predictors be $\mathcal I_{\T}=\{t_1=1,t_2=10, t_3=20, t_4=30, t_5=40, t_6=50\}$. Let $\hat d=\min(n,p)$ if there is no spiked eigenvalue, and the implementation issue regarding $c_{n_1}$ and $c_{n_2}$ is provided in Supplementary Material S.2.

\begin{example}
In this example, we consider the classic linear model.
\begin{equation}
y = x_{t_1} + x_{t_2} + x_{t_3} + x_{t_4} + x_{t_5} + x_{t_6} + \sigma\epsilon, \label{sim1-1}
\end{equation}
where $\epsilon$ is the stochastic error that follows a standard normal distribution. For setting (1) we let $\pLd=\text{diag}(80+ \lceil p/\sqrt{n} \rceil, 79+ \lceil p/\sqrt{n} \rceil,\ldots, \lceil p/\sqrt{n} \rceil,1,\ldots,1)$, where $\lceil p/\sqrt{n} \rceil$ denotes the minimum integer that is larger than $p/\sqrt{n}$. Thus, there are $81$ spiked eigenvalues for model (\ref{sim1-1}). By specifying $n, p$ and $\sigma$ at different values, we have the following five scenarios.
\begin{align*}
& \textbf{Scenario 1.1: } n = 500, p = 700, \sigma = 1; \quad 
& \textbf{Scenario 1.2: } n = 500, p = 1500, \sigma = 1; \\
& \textbf{Scenario 1.3: } n = 500, p = 1500, \sigma = 1.5; \quad
& \textbf{Scenario 1.4: } n = 500, p = 2000, \sigma = 1; \\
& \textbf{Scenario 1.5: } n = 300, p = 1000, \sigma = 1.
\end{align*}
For setting (2), we let $n, p, \rho$ and $\sigma$ be the following values.
\begin{align*}
& \textbf{Scenario 1.6: } n = 500, p = 100, \rho = 0.5, \sigma = 1;
& \textbf{Scenario 1.7: } n = 500, p = 1000, \rho = 0.5, \sigma = 1; \\
& \textbf{Scenario 1.8: } n = 500, p = 1000, \rho = 0.5, \sigma = 1.5;
& \textbf{Scenario 1.9: } n = 500, p = 1500, \rho = 0.5, \sigma = 1; \\
& \textbf{Scenario 1.10: } n = 300, p = 1000, \rho= 0.3, \sigma = 1.
\end{align*}
For each scenario, we generated $100$ datasets and applied SIRS, DC-SIS and WLS to each dataset. The means and standard deviations of the resulting FP, FN, $\mathcal M$ values and CPU time are reported in Table \ref{tb:sim1-1}.
Since there exist $6$ true predictors and $(p-6)$ irrelevant variables, the FP and FN range from $0$ to $(p-6)$ and $0$ to $6$ respectively, with smaller values indicating better performances in variable screening. 

In setting (1), WLS outperforms other methods in terms of FN and minimum model size $\mathcal M$ in all scenarios even when the variance of noise increases (scenario 1.3) and the sample size becomes smaller (scenario 1.5), and its performance keeps up with diverging $p$ (scenarios 1.1-1.4). DC-SIS and SIRS tend to miss one to three predictors on average and have larger $\mathcal M$ values as $p$ diverges or as $n$ gets smaller (scenarios 1.4-1.5). Moreover, it only takes WLS seconds to perform variable screening, much efficient than the other two methods.

In setting (2), WLS and DC-SIS successfully select all true predictors (FN = $0.00$), while keeping falsely selected predictors to a manageable size. SIRS has slightly larger FN values when there exist moderate correlations between predictors in the $p>n$ scenarios. The average minimum model size $\mathcal M$ of WLS and DC-SIS are around $6$, indicating that the true predictors have higher rankings than redundant predictors. When the variance of the noise and the number of predictors gets larger or the sample size gets smaller, the $\mathcal M$ values of WLS is slightly larger than that of DC-SIS. It is expected since there are no spiked eigenvalues that exist in this model, and thus the signals are not large enough to be detected. Furthermore, the computation time of WLS also increases. Since the number of singular vectors used to calculate WLS can be as large as $n$, it takes extra time to perform the calculation. 

\renewcommand{\arraystretch}{0.65}
\begin{center}
\captionof{table}{Performance comparison in example 1.}
\label{tb:sim1-1}
\begin{tabular*}{\textwidth}{@{\extracolsep{\fill}} c l cccc}
\hline\hline
Setting (1) & Method & FP & FN & $\mathcal M$ & Time (s) \\ \hline
\multirow{3}{*}{Scenario 1.1}
& SIRS & 74.00 (0.00) & 0.00 (0.00) & 58.65 (0.89) &  7.53 (0.56) \\ 
& DC-SIS & 74.00 (0.00) & 0.00 (0.00) & 14.58 (0.50) & 23.98 (1.29) \\ 
& WLS & 28.95 (0.72) & 0.00 (0.00) & 12.60 (0.57) &  0.26 (0.02) \\  \hline
\multirow{3}{*}{Scenario 1.2}
& SIRS & 74.00 (0.00) & 0.00 (0.00) & 31.34 (0.57) & 16.44 (1.71) \\ 
& DC-SIS & 74.00 (0.00) & 0.00 (0.00) & 27.53 (0.83) & 41.02 (2.89) \\ 
& WLS & 72.87 (0.84) & 0.00 (0.00) &  8.16 (0.58) &  0.43 (0.05) \\ \hline
\multirow{3}{*}{Scenario 1.3}
& SIRS & 74.00 (0.00) & 0.00 (0.00) & 31.63 (0.88) & 11.51 (0.17) \\
& DC-SIS & 74.00 (0.00) & 0.00 (0.00) & 27.35 (1.12) & 44.32 (1.42) \\
& WLS & 72.90 (0.89) & 0.00 (0.00) &  8.33 (0.80) &  1.49 (0.02) \\ \hline
\multirow{3}{*}{Scenario 1.4}
& SIRS & 75.75 (0.44) & 1.75 (0.44) & 179.48 (2.46) & 21.73 (2.13) \\ 
& DC-SIS & 75.00 (0.00) & 1.00 (0.00) &  98.67 (1.35) & 54.64 (3.82) \\ 
& WLS & 97.33 (0.85) & 0.00 (0.00) &  31.28 (2.69) &  0.54 (0.06) \\ \hline
\multirow{3}{*}{Scenario 1.5}
& SIRS & 49.00 (0.00) & 3.00 (0.00) & 252.70 ( 3.11) & 3.37 (0.08) \\
& DC-SIS & 47.00 (0.00) & 1.00 (0.00) &  89.21 ( 1.39) & 9.32 (0.25) \\
& WLS & 52.27 (1.06) & 0.74 (0.66) &  64.65 (12.61) & 0.43 (0.01) \\
\hline \hline
Setting (2) & Method & FP & FN & $\mathcal M$ & Time (s) \\ \hline
\multirow{3}{*}{Scenario 1.6}
& SIRS & 74.00 (0.00) & 0.00 (0.00) & 9.89 (1.44) & 3.00 (0.26) \\
& DC-SIS & 74.00 (0.00) & 0.00 (0.00) & 6.00 (0.00) & 7.82 (0.67) \\
& WLS & 14.47 (1.27) & 0.00 (0.00) & 6.00 (0.00) & 0.30 (0.03) \\ \hline
\multirow{3}{*}{Scenario 1.7}
& SIRS & 74.01 (0.10) & 0.01 (0.10) & 43.95 (12.35) & 29.75 (2.51) \\
& DC-SIS & 74.00 (0.00) & 0.00 (0.00) &  6.00 ( 0.00) & 78.63 (7.26) \\
& WLS & 45.89 (1.29) & 0.00 (0.00) &  6.01 ( 0.10) & 53.74 (3.72) \\ \hline
\multirow{3}{*}{Scenario 1.8}
& SIRS & 74.29 (0.46) & 0.29 (0.46) & 68.52 (30.06) & 29.84 (2.49) \\
& DC-SIS & 74.00 (0.00) & 0.00 (0.00) &  6.07 ( 0.29) & 79.12 (7.48) \\
& WLS & 48.16 (1.29) & 0.00 (0.00) &  6.11 ( 0.40) & 53.91 (3.98) \\ \hline
\multirow{3}{*}{Scenario 1.9}
& SIRS & 74.03 (0.17) & 0.03 (0.17) & 41.65 (17.53) &  44.83 ( 3.97) \\
& DC-SIS & 74.00 (0.00) & 0.00 (0.00) &  6.00 ( 0.00) & 118.20 (10.71) \\
& WLS & 71.06 (1.37) & 0.00 (0.00) &  6.01 ( 0.10) &  80.21 ( 6.01) \\ \hline
\multirow{3}{*}{Scenario 1.10} 
&SIRS & 46.41 (0.53) & 0.41 (0.53) & 53.64 (18.16) &  3.49 (0.03) \\
& DC-SIS & 46.00 (0.00) & 0.00 (0.00) &  6.01 ( 0.10) & 10.20 (0.04) \\
&  WLS & 32.78 (1.05) & 0.00 (0.00) &  7.89 ( 2.20) &  1.46 (0.01) \\
\hline \hline
\end{tabular*}
\end{center}

\end{example}

\begin{example}
In this example, we consider the multiple index model with the following form.
\begin{equation}
y =\frac{x_{t_1} + x_{t_2} + 1.5 x_{t_3} + 1.2 x_{t_4}}{0.5 + (x_{t_5} + 1.2x_{t_6}+1)^2 } + \sigma\epsilon, \label{sim1-2}
\end{equation}
where $\epsilon$ is the stochastic error that follows a standard normal distribution. For setting (1) we let $\pLd=\text{diag}(50+ \lceil p/\sqrt{n} \rceil, 49+ \lceil p/\sqrt{n}\rceil,\ldots, \lceil p/\sqrt{n}\rceil, 1,\ldots,1)$ and $\pV$ be identity matrix. Thus, there are $51$ spiked eigenvalues for model (\ref{sim1-2}). 
By specifying $n, p$ and $\sigma$ at different values, we have the following five scenarios.
\begin{align*}
& \textbf{Scenario 2.1: } n = 1000, p = 1200, \sigma = 1; \quad 
& \textbf{Scenario 2.2: } n = 1000, p = 1500, \sigma = 1; \\
& \textbf{Scenario 2.3: } n = 1000, p = 1500, \sigma = 1.5; \quad 
& \textbf{Scenario 2.4: } n = 1000, p = 2000, \sigma = 1; \\
& \textbf{Scenario 2.5: } n = 300, p = 2000, \sigma = 1.
\end{align*}
For setting (2), we let $n, p, \rho$ and $\sigma$ be the following values.
\begin{align*}
& \textbf{Scenario 2.6: } n = 1000, p = 200, \rho = 0.5, \sigma = 1;
& \textbf{Scenario 2.7: } n = 1000, p = 2000, \rho = 0.5, \sigma = 1; \\
& \textbf{Scenario 2.8: } n = 1000, p = 2000, \rho = 0.5, \sigma = 1.5;
& \textbf{Scenario 2.9: } n = 1000, p = 2500, \rho = 0.5, \sigma = 1; \\
& \textbf{Scenario 2.10: } n = 500, p = 2000, \rho=0.3, \sigma = 1;
\end{align*}
In each scenario, we generated $100$ datasets and applied SIRS, DC-SIS and WLS to each dataset. The means and standard deviations of the resulting FP, FN, $\mathcal M$ values and CPU time are reported in Table \ref{tb:sim1-2}. 

In setting (1), WLS works better in screening redundant predictors (FP, scenarios 2.1 - 2.5) compared with SIRS, especially when the number of redundant predictors and errors of the model increase. DC-SIS misses two to four predictors on average. Notice that in this setting, $\pV$ is an identity matrix and the $p$ candidate predictors are nearly independent. This model setting favors SIRS since SIRS requires that there is not strong collinearity between the true and redundant predictors or among the true predictors themselves. Regarding the minimum model size $\mathcal M$, WLS ranks first, indicating that WLS is able to find all true predictors with the smallest model size.

In setting (2), predictors are assumed to have moderate correlations. WLS has better performances regarding FP and FN values especially when $p$ diverges. It implies that WLS is able to include all true predictors while keeping FP value to a manageable size. SIRS on average misses two predictors when there exist moderate correlations between predictors in the $p>n$ scenarios (scenarios 2.7-2.10).
WLS ranks first concerning the minimum model size $\mathcal M$. 

\renewcommand{\arraystretch}{0.6}
\begin{center}
\captionof{table}{Performance comparison in example 2.}
\label{tb:sim1-2}
\begin{tabular*}{\textwidth}{@{\extracolsep{\fill}} c l cccc}
\hline\hline
Setting (1) & Method & FP & FN & $\mathcal M$ & Time (s) \\ \hline
\multirow{3}{*}{Scenario 2.1}
& SIRS & 138.00 (0.00) & 0.00 (0.00) &  38.98 (  8.44) & 111.21 (10.29) \\
& DC-SIS & 140.26 (0.48) & 2.26 (0.48) & 664.63 (100.18) & 386.04 (41.13) \\
& WLS &  42.94 (1.29) & 0.04 (0.20) &  36.01 ( 10.56) &   8.94 ( 1.22) \\ \hline
\multirow{3}{*}{Scenario 2.2}
& SIRS & 138.00 (0.00) & 0.00 (0.00) &   44.22 (  5.13) & 137.89 (12.00) \\
& DC-SIS & 140.17 (0.43) & 2.17 (0.43) & 1190.07 (218.93) & 476.11 (45.02) \\
& WLS &  44.13 (0.87) & 0.03 (0.17) &   36.21 ( 10.36) &  11.29 ( 1.61) \\ \hline
\multirow{3}{*}{Scenario 2.3}
& SIRS & 138.00 (0.00) & 0.00 (0.00) &   44.57 (  5.06) &  36.41 (0.55) \\
& DC-SIS & 140.39 (0.65) & 2.39 (0.65) & 1190.12 (246.29) & 210.17 (7.13) \\
& WLS &  44.21 (0.83) & 0.00 (0.00) &   36.20 (  9.98) &   5.59 (0.10) \\ \hline
\multirow{3}{*}{Scenario 2.4}
& SIRS & 138.00 (0.00) & 0.00 (0.00) &   40.44 (  8.20) & 184.18 (16.60) \\
& DC-SIS & 140.76 (0.43) & 2.76 (0.43) & 1490.83 (196.61) & 636.14 (64.02) \\
& WLS &  44.73 (0.51) & 0.03 (0.17) &   38.20 ( 10.32) &  16.69 ( 2.74) \\ \hline
\multirow{3}{*}{Scenario 2.5}
&SIRS & 46.00 (0.00) & 0.00 (0.00) &   46.97 (  3.67) &  7.03 (0.22) \\
& DC-SIS & 50.86 (0.35) & 4.86 (0.35) & 1876.98 (125.51) & 20.18 (0.37) \\
& WLS & 44.92 (0.27) & 0.00 (0.00) &   42.39 (  7.49) &  0.66 (0.04) \\
\hline \hline
Setting (2) & Method & FP & FN & $\mathcal M$ & Time (s) \\ \hline
\multirow{3}{*}{Scenario 2.6}
& SIRS & 138.00 (0.00) & 0.00 (0.00) & 30.36 (7.08) & 19.40 (1.64) \\
& DC-SIS & 138.00 (0.00) & 0.00 (0.00) & 12.77 (1.78) & 62.92 (5.38) \\
& WLS &  31.83 (1.98) & 0.00 (0.00) &  6.14 (0.78) &  2.10 (0.16) \\ \hline
\multirow{3}{*}{Scenario 2.7}
& SIRS & 139.96 (0.20) & 1.96 (0.20) & 483.44 (136.12) & 193.92 (16.40) \\
& DC-SIS & 138.00 (0.00) & 0.00 (0.00) &  14.33 (  1.60) & 627.38 (53.86) \\
& WLS &  89.48 (1.76) & 0.00 (0.00) &   7.04 (  1.34) & 427.98 (30.26) \\ \hline
\multirow{3}{*}{Scenario 2.8}
& SIRS & 140.00 (0.00) & 2.00 (0.00) & 806.54 (225.90) & 193.23 (15.74) \\
& DC-SIS & 138.00 (0.00) & 0.00 (0.00) &  28.48 ( 15.11) & 621.74 (50.78) \\
& WLS &  94.83 (1.80) & 0.01 (0.10) &  19.45 ( 16.68) & 429.94 (31.38) \\ \hline
\multirow{3}{*}{Scenario 2.9}
& SIRS & 139.98 (0.14) & 1.98 (0.14) & 575.28 (183.56) & 242.22 (20.61) \\
& DC-SIS & 138.00 (0.00) & 0.00 (0.00) &  14.98 (  2.59) & 787.96 (66.91) \\
& WLS & 115.94 (1.85) & 0.00 (0.00) &  11.17 (  8.93) & 536.32 (38.58) \\ \hline
\multirow{3}{*}{Scenario 2.10}
& SIRS & 76.53 (0.50) & 2.53 (0.50) & 988.85 (250.46) & 16.69 (1.07) \\
&  DC-SIS & 74.05 (0.22) & 0.05 (0.22) &  31.42 ( 22.88) & 65.89 (4.28) \\
&  WLS & 65.18 (1.28) & 0.05 (0.22) &  27.72 ( 20.01) &  8.31 (0.39) \\
\hline \hline
\end{tabular*}
\end{center}

\end{example}
\vspace{-10pt}
\begin{example}
In previous examples, the true predictors affect the mean response. In this example, we consider the heteroscedastic model of the following form.
\begin{equation}
 y = \frac{\sigma\epsilon}{1 + 1.2x_{t_1} + x_{t_2} + x_{t_3} + 1.5x_{t_4} + x_{t_5} + x_{t_6}}, \label{sim1-3}
\end{equation}
where $\epsilon$ is the stochastic error that follows a standard normal distribution. For setting (1) we let $\pLd=\text{diag}(50+ \lceil p/\sqrt{n} \rceil, 49+ \lceil p/\sqrt{n} \rceil,\ldots, \lceil p/\sqrt{n}\rceil, 1,\ldots,1)$ and $\pV$ be identity matrix. 
By specifying $n, p$ and $\sigma$ at different values, we have the following scenarios.
\begin{align*}
& \textbf{Scenario 3.1: } n = 1000, p = 1200, \sigma = 1; \quad 
& \textbf{Scenario 3.2: } n = 1000, p = 1500, \sigma = 1; \\
& \textbf{Scenario 3.3: } n = 1000, p = 2000, \sigma = 1; \quad
& \textbf{Scenario 3.4: } n = 300, p = 2000, \sigma = 1.
\end{align*}
\vspace{-10pt}
For setting (2), we let $n, p, \rho$ and $\sigma$ be the following values.
\begin{align*}
& \textbf{Scenario 3.5: } n = 1000, p = 200, \rho = 0.3, \sigma = 1;
& \textbf{Scenario 3.6: } n = 1000, p = 2000, \rho = 0.1, \sigma = 1; \\
& \textbf{Scenario 3.7: } n = 1000, p = 2500, \rho = 0.1, \sigma = 1;
& \textbf{Scenario 3.8: } n = 500, p = 2000, \rho=0.1, \sigma = 1.
\end{align*}
In each scenario, we generated $100$ datasets and applied SIRS, DC-SIS and WLS to each dataset. The means and standard deviations of the resulting FP, FN, $\mathcal M$ values and CPU time are reported in Table \ref{tb:sim1-3}. 

In setting (1), by investigating FP and FN values, we find that both WLS and SIRS enjoy good performance for this model and correctly recover all true predictors with large probabilities. This model setting also favors SIRS and thus it works reasonably well. DC-SIS misses five predictors on average, as the minimum distance correlation of active predictors are too small to be detected. Regarding the minimum model size $\mathcal M$, WLS and SIRS have comparable performance and are stable under various scenarios.

In setting (2), WLS still enjoys good performance in heteroscedastic model when there is no spiked eigenvalues. As $p$ diverges (scenarios 3.6 and 3.7), WLS attains the lowest FP and FN values, while DC-SIS and SIRS on average miss two to five predictors. Regarding the average minimum model size $\mathcal M$, WLS outperforms SIRS and DC-SIS in all scenarios.

\renewcommand{\arraystretch}{0.6}
\begin{center}
\captionof{table}{Performance comparison in example 3.}
\label{tb:sim1-3}
\begin{tabular*}{\textwidth}{@{\extracolsep{\fill}} c l cccc}
\hline\hline
Setting (1) & Method & FP & FN & $\mathcal M$ & Time (s) \\ \hline
\multirow{3}{*}{Scenario 3.1}
& SIRS & 138.00 (0.00) & 0.00 (0.00) &   40.45 (  6.87) & 110.27 ( 9.71) \\
& DC-SIS & 143.81 (0.51) & 5.81 (0.51) & 1015.60 (149.40) & 379.41 (38.36) \\
& WLS &  43.66 (1.10) & 0.22 (0.42) &   45.23 (  5.22) &   8.95 ( 1.32) \\  \hline
\multirow{3}{*}{Scenario 3.2}
& SIRS & 138.00 (0.00) & 0.00 (0.00) &   45.19 (  4.91) & 137.50 (11.90) \\
& DC-SIS & 143.39 (0.85) & 5.39 (0.85) & 1303.80 (198.85) & 474.86 (46.08) \\
& WLS &  44.32 (0.82) & 0.07 (0.26) &   44.73 (  5.64) &  11.22 ( 1.53) \\ \hline
\multirow{3}{*}{Scenario 3.3}
& SIRS & 138.00 (0.00) & 0.00 (0.00) &   43.67 (  6.52) & 183.81 (14.85) \\
& DC-SIS & 143.34 (0.54) & 5.34 (0.54) & 1734.90 (225.29) & 627.75 (54.28) \\
& WLS &  44.73 (0.69) & 0.03 (0.17) &   43.93 (  5.89) &  16.16 ( 2.20) \\ \hline
\multirow{3}{*}{Scenario 3.4}
& SIRS & 46.00 (0.00) & 0.00 (0.00) &   46.20 (  4.38) &  7.03 (0.23) \\
&  DC-SIS & 51.86 (0.35) & 5.86 (0.35) & 1724.04 (251.56) & 20.17 (0.39) \\
&  WLS & 44.95 (0.22) & 0.04 (0.20) &   45.17 (  6.76) &  0.66 (0.04) \\
\hline \hline
Setting (2) & Method & FP & FN & $\mathcal M$ & Time (s) \\ \hline
\multirow{3}{*}{Scenario 3.5}
& SIRS & 138.00 (0.00) & 0.00 (0.00) & 71.22 (16.63) & 19.51 (1.61) \\
& DC-SIS & 138.05 (0.22) & 0.05 (0.22) & 65.98 (39.43) & 63.30 (5.94) \\
& WLS &  51.38 (2.02) & 0.25 (0.44) & 46.02 (40.60) &  2.11 (0.17) \\ \hline
\multirow{3}{*}{Scenario 3.6}
& SIRS & 141.94 (0.65) & 3.94 (0.65) & 902.59 (183.33) & 194.20 (15.59) \\
& DC-SIS & 141.38 (1.15) & 3.38 (1.15) & 728.99 (306.22) & 631.71 (54.10) \\
& WLS & 110.18 (1.78) & 1.28 (0.96) & 310.62 (251.74) & 429.31 (29.71) \\ \hline
\multirow{3}{*}{Scenario 3.7}
& SIRS & 143.22 (0.73) & 5.22 (0.73) & 1134.49 (275.10) & 243.07 (20.39) \\
& DC-SIS & 139.71 (0.71) & 1.71 (0.71) &  897.10 (519.68) & 782.91 (68.00) \\
& WLS & 138.29 (1.91) & 1.31 (0.85) &  679.12 (548.77) & 536.98 (38.92) \\ \hline
\multirow{3}{*}{Scenario 3.8}
&SIRS & 79.58 (0.55) & 5.58 (0.55) & 1144.52 (236.43) & 16.50 (0.28) \\
&  DC-SIS & 78.41 (0.71) & 4.41 (0.71) & 1345.58 (356.08) & 68.58 (4.02) \\
&  WLS & 70.68 (1.61) & 2.91 (0.75) &  965.09 (500.01) &  8.58 (0.78) \\
\hline \hline
\end{tabular*}
\end{center}

\end{example}

To conclude, SIRS and DC-SIS, as extensions of SIS, can be applied to a wide range of parametric and semi-parametric models and are particularly appealing for variable screening when the number of candidate predictors exceeds the sample size. However,
SIRS requires there to be no strong collinearity between the true and redundant predictors or among the true predictors themselves. SIRS thus may fail to identify the true predictor that is correlated with redundant predictors. As illustrated in example 2-3 setting (2), when there exists moderate correlations between predictors, SIRS fails to identify two to five true predictors on average with a diverging $p$. While DC-SIS may also fail to identify some important predictors that have small marginal distance correlations with the response (example 2-3 setting (1)). For WLS screening method, simulation studies show that it is a robust variable screening method under various scenarios, even when the covariance of the predictors does not have spiked eigenvalues (example 1-3 setting (2)). 
\section{Weighted leverage score for cancer biomarker detection}\label{sec:app}

Cancer, characterized by uncontrolled abnormal cell growth and invasion, has gradually become the primary cause of death across the world. According to the National Cancer Institute, more than $1.68$ million new cases of cancer will be diagnosed in the United States, and nearly $0.6$ million people would die from the disease. Although national expenditures for cancer care and cancer research are tremendous, cancer survival rates still tend to be poor due to late diagnosis. Therefore, an early and accurate detection of cancer is of primary importance. 

With the recent advancement in next generation sequencing technology, accurate detection of cancer becomes possible and holds tremendous promise. It has been shown that many cancers have altered messenger RNA (mRNA) metabolism \citep{wu2015cancer}. In tumor cells, there exists aberrant mRNA processing, nuclear export, and translations, which may lead to the loss of function of some tumor suppressors \citep{pandolfi2004aberrant, siddiqui2012mrna, wu2015cancer}. One typical inference thus is to find the tumor-related marker genes that can discriminate cancer patients from normal and early-stage cancer from late-stage. This can be achieved using the variable selection approach under the classification or regression model.
However, in a typical biomarker detection, the number of identified non-invasive/invasive cancer subjects is only in the hundreds, while the number of candidate marker genes is usually in the tens of thousands. Most existing statistical methods are inapplicable in this notorious ``small $n$, ultra-large $p$'' setting. There is a further layer of complications when there exists a nonlinear relationship between gene expression levels and cell types within tissue sections, because the nonlinear models are more susceptible to the curse of dimensionality. Effective variable selection methods for nonlinear models thus are even more critical than that for linear models in identifying marker genes for the early cancer detection.

To identify marker genes, we applied the WLS screening approach to analyze the breast cancer spatial transcriptomics data \citep{staahl2016visualization}. Spatial transcriptomics is a recent sequencing strategy that allows the quantification of gene expression with spatial resolution in individual tissue sections. Standard RNA-seq technique produces an averaged transcriptome, while spatial transcriptomics simultaneously sequences different locations of a breast cancer tissue section, including normal, cancer, and invasive cancer areas.
This strategy provides gene expression data with less noise. In this experiment, $518$ locations on two histological sections that from a breast cancer biopsy were sequenced, among which $64$ were identified as invasive cancer areas, $73$ were identified as non-invasive cancer areas, and $381$ were identified as non-cancer areas. Those locations were identified based on morphological criteria \citep{staahl2016visualization}. In each location, expressions of $3572$ genes were quantified. To build a predictive model as illustrated in (\ref{eq2}), we treat location labels as the response variable and the expression values of $3572$ genes as predictors. More precisely, the response is a vector with $518$ entries and the data matrix is a $518 \times 3572$ matrix with $(i,j)$th entry representing the expression of gene $j$ at area $i$. 

\begin{figure}[H]
  \centering
  \includegraphics[width=1\textwidth]{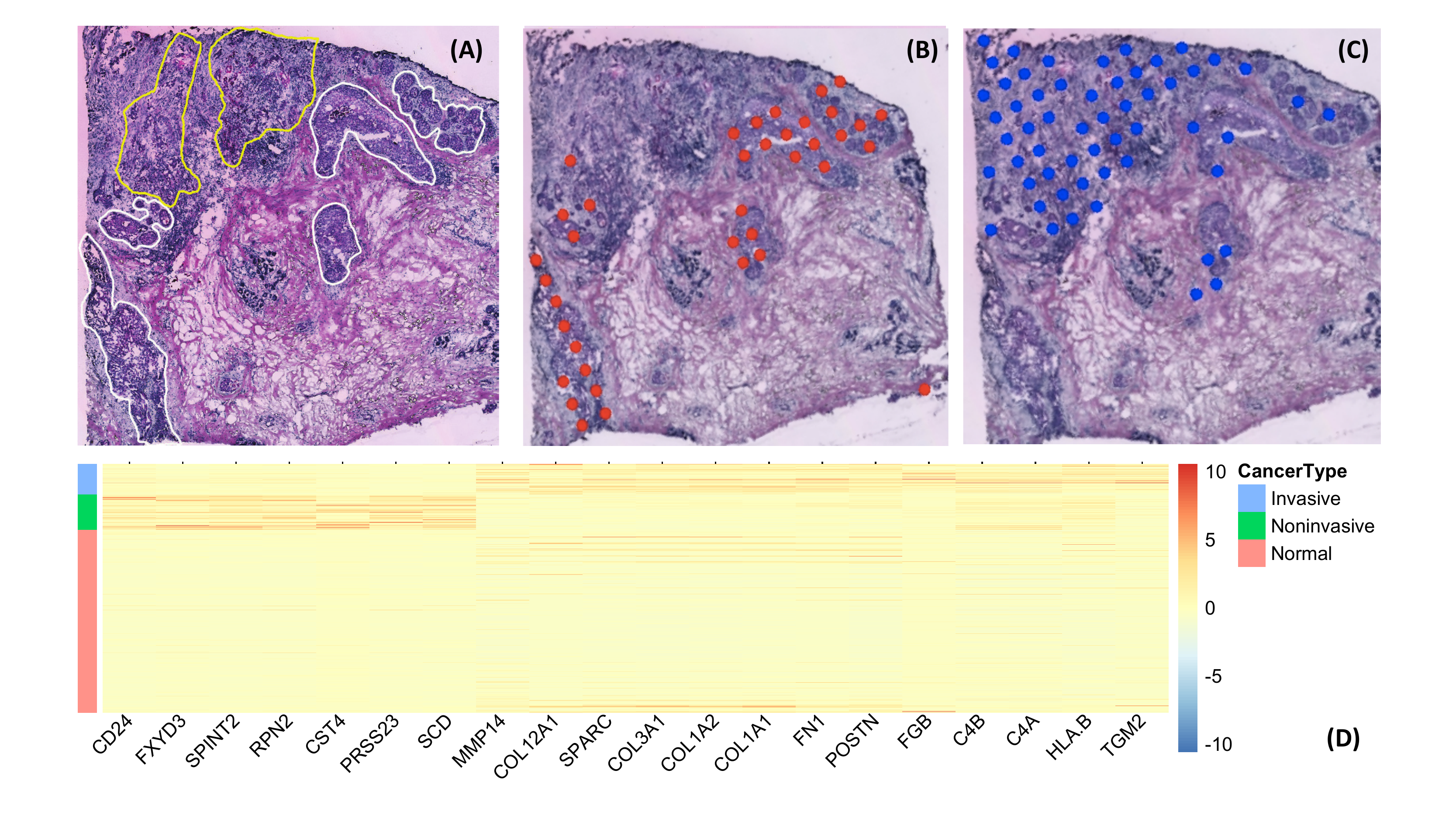}
  \caption{(A) is one histological section from breast cancer biopsy with two areas of invasive ductal cancer (yellow line) and four areas of ductal cancer in situ (white line). Other areas are non-cancer areas. The image is obtained from \cite{staahl2016visualization}. (B) shows the areas where genes \textit{PRSS23} and \textit{SCD} were highly expressed. (C) shows the areas where genes \textit{FGB}, \textit{TGM2} and \textit{FN1} were highly expressed. (D) is a heatmap of expressions of genes selected by WLS. For the ease of presentation, we only showed the first $20$ genes.}
  \label{fig:realdata}
\end{figure}

We applied WLS to this data set and identified $225$ genes that were differentially expressed between invasive, noninvasive and normal areas. The revealed expression patterns show a remarkable spatial difference in gene expression profiles between areas of cancer (Figure \ref{fig:realdata} (B) and (C)). For example, genes \textit{PRSS23} and \textit{SCD} were highly expressed in non-invasive cancer areas (Figure \ref{fig:realdata} (B)), and genes \textit{FGB}, \textit{TGM2} and \textit{FN1} were highly expressed in invasive cancer areas (Figure \ref{fig:realdata} (C)). 
To understand the biological processes that those genes were involved, we also annotated the functions of them using Gene Ontology Consortium. The $225$ genes were enriched in $47$ functional classes. In particular, $41$ genes were involved in regulation of cell death, and $38$ of them were involved in regulation of apoptotic process, one of the most important cancer hallmarks. It was also of interest to note that the three genes: \textit{FGB} (fibrinogen beta chain), \textit{TGM2} (transglutaminase 2), and \textit{FN1} (fibronectin 1) that were over-expressed in invasive cancer areas were involved in apoptotic process. The increased expression and activation of tissue transglutaminase (\textit{TGM2}) often occurred in response to the stimuli that promote cell differentiation and apoptosis, which further contributes to its oncogenic potential in breast cancer cells \citep{antonyak2004augmentation}. The expression of \textit{FN1} was regulated by micro (mi)RNA-206 who was demonstrated to be associated with metastatic cancer types, including breast cancer \citep{liu2015identification, kondo2008mir, adams2009argonaute}. \textit{FN1} gene itself was also found to be a key regulator in breast cancer development \citep{liu2015identification} and be correlated with the drug resistance of cancer cells \citep{mutlu2012differential}. Other genes were involved in pathways that may contribute to cancer development such as blood vessel development, and pathways that related to immune system such as neutrophil mediated immunity and cell activation during immune response. \cite{staahl2016visualization} performed the differential expression analysis on this dataset using the DESeq2 tool \citep{love2014moderated}, a negative binomial model-based hypothesis testing method. Several genes (\emph{IGFBP5, MUCL1, PIP, FN1, POSTN, SPARC, MMP14}) were highlighted in the paper and were overlapped with the feature genes identified by the WLS method. Moreover, WLS identified other genes that were enriched in the apoptotic process and were in need for further investigation. Since WLS is a model-free variable screening method, it is able to detect predictors when the relationship between them and the response is beyond linear.  

We also applied the methods SIRS and DC-SIS to this dataset. 
The SIRS method detected $82$ feature genes, among which $17$ were enriched in the regulation of cell death and the regulation of apoptotic process. The DC-SIS method also detected $82$ feature genes involved in the regulation of cell growth and pathways that may contribute to cell development. To evaluate the prediction accuracy of the WLS method, we further trained random forests to predict sample's identity using the identified feature genes. The 10-fold cross-validation results were reported in Table \ref{tb:real_acc}. In terms of the prediction accuracy, the WLS method outperformed other methods.
\renewcommand{\arraystretch}{0.6}
\begin{center}
\captionof{table}{Prediction Accuracy}
\label{tb:real_acc}
\begin{tabular*}{\textwidth}{@{\extracolsep{\fill}} l cccc}
  \hline\hline
Method & Invasive Group & Noninvasive Group & Normal Group & Overall \\
  \hline
SIRS & 0.4622 & 0.7879 & 0.9609 & 0.8687 \\
DC-SIS & 0.4288 & 0.8137 & 0.9659 & 0.8745 \\
WLS & 0.4622 & 0.8303 & 0.9717 & 0.8842 \\
   \hline\hline
\end{tabular*}
\end{center}

\section{Discussion}\label{sec:dis}

Leverage score has long been used for model diagnostics in linear regression. Recently, leverage score has been shown to be a powerful tool for big data analytics. Subsamples that randomly selected according to the leverage scores are good surrogates of the full sample in estimating linear regression models. Thus it is extensively used to overcome the computational challenges that arise from analyzing a massive number of samples. Despite the promising results of leverage score sampling in reducing big sample size $n$, it remains elusive how it can be used to reduce the dimensionality when $p$ is large.

The WLS screening method generalizes the recent work \citep{ma2014statistical,ma2015leveraging} on leverage score based sampling to predictors screening. The proposed screening procedure has a novel contribution to the literature of variable screening for high-dimensional regression analysis. First, it is developed under the SDR framework and does not impose any assumption on the relationship between the response and predictors. Second, compared with existing variable screening methods under the SDR framework, it is a more potent tool in real applications since there is no need to pre-specify the number of linear combinations $k$. Third, it can handle the data with a large number of candidate predictors, especially when $p >> n$, which is highly desirable for the high-dimensional setting. Finally, WLS generalizes the concept of leverage score in linear models for sub-sampling to variable screening in nonparametric models. It is derived based on both the right and left leverage scores and consistently evaluates the importance of predictors. Thus it enjoys an excellent computational and theoretical advantage. 

As a trade-off, the WLS screening procedure imposes a few assumptions on the distribution of the predictors, of which the design condition is fundamental and crucial. It requires that the predictors are from a non-degenerate elliptically symmetric distribution. For a consistent estimate of $\omega_j$ in the scenario of $p/(\sqrt{n}\lambda_d^2) \rightarrow 0$ when $n, p$ and $\lambda_d$ go to infinity, the sub-Gaussian distribution is imposed to predictors. These assumptions ensure the ranking consistency of WLS for variable screening in high-dimensional data. The WLS also depends on a fixed slicing scheme, which is more of a technical issue. For the slicing scheme, the allowed number of observations within each slice is as close to each other as possible, while the range of each slice may vary. When choosing the number of slices $h$, we recommend to have at least $10$ observations within each slice, and a larger number of slices is preferred to ensure selection consistency \citep{zhong2012correlation}. As discussed in \cite{li1991sliced}, inappropriate choices of $h$ may result in a slower convergence rate but would not lead to a significant differences in the behaviors of the output. Thus, instead of making the mathematical formulation of the WLS method more complicated, we choose to focus on the fixed slicing scheme for practical considerations.

The WLS screening approach provides a rich and flexible framework to address the curse of dimensionality in regression. We believe that the results from this project will make significant theoretical and methodological contributions to the study of general index models and variable screening algorithms, and have a broad and important impact on applications in many areas. To facilitate the method development in this direction, we implemented the WLS screening algorithm using programming language R, and the source code can be downloaded from \href{https://github.com/yiwenstat/wls.git}{Github}.

\section*{Acknowledgment}
The authors would like to acknowledge the support from the U.S. National Science Foundation under grants DMS-1903226 and DMS-1925066, and the U.S. National Institute of Health under grant 5R01GM113242.


\bibliographystyle{chicago}
\bibliography{reference/dr,reference/ref, reference/response}

\begin{thebibliography}{}

\bibitem[\protect\citeauthoryear{Adams, Claffey, and White}{Adams
  et~al.}{2009}]{adams2009argonaute}
Adams, B.~D., K.~P. Claffey, and B.~A. White (2009).
\newblock Argonaute-2 expression is regulated by epidermal growth factor
  receptor and mitogen-activated protein kinase signaling and correlates with a
  transformed phenotype in breast cancer cells.
\newblock {\em Endocrinology\/}~{\em 150\/}(1), 14--23.

\bibitem[\protect\citeauthoryear{Allen-Zhu and Li}{Allen-Zhu and
  Li}{2016}]{allen2016lazysvd}
Allen-Zhu, Z. and Y.~Li (2016).
\newblock Lazy{SVD}: Even faster {SVD} decomposition yet without agonizing
  pain.
\newblock In {\em Advances in Neural Information Processing Systems}, pp.\
  974--982.

\bibitem[\protect\citeauthoryear{Antonyak, Miller, Jansen, Boehm, Balkman,
  Wakshlag, Page, and Cerione}{Antonyak
  et~al.}{2004}]{antonyak2004augmentation}
Antonyak, M.~A., A.~M. Miller, J.~M. Jansen, J.~E. Boehm, C.~E. Balkman, J.~J.
  Wakshlag, R.~L. Page, and R.~A. Cerione (2004).
\newblock Augmentation of tissue transglutaminase expression and activation by
  epidermal growth factor inhibit doxorubicin-induced apoptosis in human breast
  cancer cells.
\newblock {\em Journal of Biological Chemistry\/}~{\em 279\/}(40),
  41461--41467.

\bibitem[\protect\citeauthoryear{Breiman}{Breiman}{1995}]{breiman1995better}
Breiman, L. (1995).
\newblock Better subset regression using the nonnegative garrote.
\newblock {\em Technometrics\/}~{\em 37\/}(4), 373--384.

\bibitem[\protect\citeauthoryear{Chen and Chen}{Chen and
  Chen}{2008}]{chen2008extended}
Chen, J. and Z.~Chen (2008).
\newblock Extended bayesian information criteria for model selection with large
  model spaces.
\newblock {\em Biometrika\/}~{\em 95\/}(3), 759--771.

\bibitem[\protect\citeauthoryear{Cook}{Cook}{1995}]{cook1995introduction}
Cook, D. (1995).
\newblock An introduction to regression graphics.
\newblock {\em Journal of the American Statistical Association\/}~{\em
  90\/}(431), 1126--1128.

\bibitem[\protect\citeauthoryear{Cook}{Cook}{1996}]{cook1996graphics}
Cook, R.~D. (1996).
\newblock Graphics for regressions with a binary response.
\newblock {\em Journal of the American Statistical Association\/}~{\em
  91\/}(435), 983--992.

\bibitem[\protect\citeauthoryear{Cook}{Cook}{1998}]{cook1998principal}
Cook, R.~D. (1998).
\newblock Principal hessian directions revisited.
\newblock {\em Journal of the American Statistical Association\/}~{\em
  93\/}(441), 84--94.

\bibitem[\protect\citeauthoryear{Cook et~al.}{Cook
  et~al.}{2004}]{cook2004testing}
Cook, R.~D. et~al. (2004).
\newblock Testing predictor contributions in sufficient dimension reduction.
\newblock {\em The Annals of Statistics\/}~{\em 32\/}(3), 1062--1092.

\bibitem[\protect\citeauthoryear{Dasgupta, Drineas, Harb, Josifovski, and
  Mahoney}{Dasgupta et~al.}{2007}]{dasgupta2007feature}
Dasgupta, A., P.~Drineas, B.~Harb, V.~Josifovski, and M.~W. Mahoney (2007).
\newblock Feature selection methods for text classification.
\newblock In {\em Proceedings of the 13th ACM SIGKDD International Conference
  on Knowledge Discovery and Data Mining}, pp.\  230--239. ACM.

\bibitem[\protect\citeauthoryear{Drineas, Kannan, and Mahoney}{Drineas
  et~al.}{2006}]{drineas2006fast}
Drineas, P., R.~Kannan, and M.~W. Mahoney (2006).
\newblock Fast monte carlo algorithms for matrices {III}: Computing a
  compressed approximate matrix decomposition.
\newblock {\em SIAM Journal on Computing\/}~{\em 36\/}(1), 184--206.

\bibitem[\protect\citeauthoryear{Drineas, Mahoney, and Muthukrishnan}{Drineas
  et~al.}{2008}]{drineas2008relative}
Drineas, P., M.~W. Mahoney, and S.~Muthukrishnan (2008).
\newblock Relative-error {CUR} matrix decompositions.
\newblock {\em SIAM Journal on Matrix Analysis and Applications\/}~{\em
  30\/}(2), 844--881.

\bibitem[\protect\citeauthoryear{Duan and Li}{Duan and
  Li}{1991}]{duan1991slicing}
Duan, N. and K.-C. Li (1991).
\newblock Slicing regression: a link-free regression method.
\newblock {\em The Annals of Statistics\/}~{\em 19\/}(2), 505--530.

\bibitem[\protect\citeauthoryear{Efroymson}{Efroymson}{1960}]{efroymson1960multiple}
Efroymson, M. (1960).
\newblock Multiple regression analysis.
\newblock {\em Mathematical Methods for Digital Computers\/}~{\em 1}, 191--203.

\bibitem[\protect\citeauthoryear{Fan, Feng, and Song}{Fan
  et~al.}{2011}]{fan2011nonparametric}
Fan, J., Y.~Feng, and R.~Song (2011).
\newblock Nonparametric independence screening in sparse ultra-high-dimensional
  additive models.
\newblock {\em Journal of the American Statistical Association\/}~{\em
  106\/}(494), 544--557.

\bibitem[\protect\citeauthoryear{Fan and Li}{Fan and
  Li}{2001}]{fan2001variable}
Fan, J. and R.~Li (2001).
\newblock Variable selection via nonconcave penalized likelihood and its oracle
  properties.
\newblock {\em Journal of the American statistical Association\/}~{\em
  96\/}(456), 1348--1360.

\bibitem[\protect\citeauthoryear{Fan and Lv}{Fan and Lv}{2008}]{fan2008sure}
Fan, J. and J.~Lv (2008).
\newblock Sure independence screening for ultrahigh dimensional feature space.
\newblock {\em Journal of the Royal Statistical Society: Series B (Statistical
  Methodology)\/}~{\em 70\/}(5), 849--911.

\bibitem[\protect\citeauthoryear{Fan and Lv}{Fan and
  Lv}{2010}]{fan2010selective}
Fan, J. and J.~Lv (2010).
\newblock A selective overview of variable selection in high dimensional
  feature space.
\newblock {\em Statistica Sinica\/}~{\em 20\/}(1), 101--148.

\bibitem[\protect\citeauthoryear{Fan, Samworth, and Wu}{Fan
  et~al.}{2009}]{fan2009ultrahigh}
Fan, J., R.~Samworth, and Y.~Wu (2009).
\newblock Ultrahigh dimensional feature selection: beyond the linear model.
\newblock {\em The Journal of Machine Learning Research\/}~{\em 10},
  2013--2038.

\bibitem[\protect\citeauthoryear{Fan and Wang}{Fan and
  Wang}{2015}]{fan2015asymptotics}
Fan, J. and W.~Wang (2015).
\newblock Asymptotics of empirical eigen-structure for ultra-high dimensional
  spiked covariance model.
\newblock {\em arXiv preprint arXiv:1502.04733\/}.

\bibitem[\protect\citeauthoryear{Gallant, Rossi, and Tauchen}{Gallant
  et~al.}{1993}]{gallant1993nonlinear}
Gallant, A.~R., P.~E. Rossi, and G.~Tauchen (1993).
\newblock Nonlinear dynamic structures.
\newblock {\em Econometrica: Journal of the Econometric Society\/}~{\em
  61\/}(4), 871--907.

\bibitem[\protect\citeauthoryear{Hall and Miller}{Hall and
  Miller}{2009}]{hall2012using}
Hall, P. and H.~Miller (2009).
\newblock Using generalized correlation to effect variable selection in very
  high dimensional problems.
\newblock {\em Journal of Computational and Graphical Statistics\/}~{\em
  18\/}(3), 533--550.

\bibitem[\protect\citeauthoryear{Hall, Titterington, and Xue}{Hall
  et~al.}{2009}]{hall2009tilting}
Hall, P., D.~Titterington, and J.-H. Xue (2009).
\newblock Tilting methods for assessing the influence of components in a
  classifier.
\newblock {\em Journal of the Royal Statistical Society: Series B (Statistical
  Methodology)\/}~{\em 71\/}(4), 783--803.

\bibitem[\protect\citeauthoryear{Huang, Horowitz, and Ma}{Huang
  et~al.}{2008}]{huang2008asymptotic}
Huang, J., J.~L. Horowitz, and S.~Ma (2008).
\newblock Asymptotic properties of bridge estimators in sparse high-dimensional
  regression models.
\newblock {\em Annals of Statistics\/}~{\em 36\/}(2), 587--613.

\bibitem[\protect\citeauthoryear{Johnstone}{Johnstone}{2001}]{johnstone2001distribution}
Johnstone, I.~M. (2001).
\newblock On the distribution of the largest eigenvalue in principal components
  analysis.
\newblock {\em Annals of Statistics\/}~{\em 29\/}(2), 295--327.

\bibitem[\protect\citeauthoryear{Kondo, Toyama, Sugiura, Fujii, and
  Yamashita}{Kondo et~al.}{2008}]{kondo2008mir}
Kondo, N., T.~Toyama, H.~Sugiura, Y.~Fujii, and H.~Yamashita (2008).
\newblock {miR}-206 expression is down-regulated in estrogen receptor
  $\alpha$--positive human breast cancer.
\newblock {\em Cancer Research\/}~{\em 68\/}(13), 5004--5008.

\bibitem[\protect\citeauthoryear{Li}{Li}{1991}]{li1991sliced}
Li, K.-C. (1991).
\newblock Sliced inverse regression for dimension reduction.
\newblock {\em Journal of the American Statistical Association\/}~{\em
  86\/}(414), 316--327.

\bibitem[\protect\citeauthoryear{Li, Zhong, and Zhu}{Li
  et~al.}{2012}]{li2012feature}
Li, R., W.~Zhong, and L.~Zhu (2012).
\newblock Feature screening via distance correlation learning.
\newblock {\em Journal of the American Statistical Association\/}~{\em
  107\/}(499), 1129--1139.

\bibitem[\protect\citeauthoryear{Liu, Ma, Yang, Wu, Jiang, and Chen}{Liu
  et~al.}{2015}]{liu2015identification}
Liu, X., Y.~Ma, W.~Yang, X.~Wu, L.~Jiang, and X.~Chen (2015).
\newblock Identification of therapeutic targets for breast cancer using
  biological informatics methods.
\newblock {\em Molecular Medicine Reports\/}~{\em 12\/}(2), 1789--1795.

\bibitem[\protect\citeauthoryear{Love, Huber, and Anders}{Love
  et~al.}{2014}]{love2014moderated}
Love, M.~I., W.~Huber, and S.~Anders (2014).
\newblock Moderated estimation of fold change and dispersion for {RNA}-seq data
  with {DESeq2}.
\newblock {\em Genome Biology\/}~{\em 15\/}(12), 550.

\bibitem[\protect\citeauthoryear{Ma, Mahoney, and Yu}{Ma
  et~al.}{2014}]{ma2014statistical}
Ma, P., M.~Mahoney, and B.~Yu (2014).
\newblock A statistical perspective on algorithmic leveraging.
\newblock In {\em International Conference on Machine Learning}, pp.\  91--99.

\bibitem[\protect\citeauthoryear{Ma and Sun}{Ma and
  Sun}{2015}]{ma2015leveraging}
Ma, P. and X.~Sun (2015).
\newblock Leveraging for big data regression.
\newblock {\em Wiley Interdisciplinary Reviews: Computational
  Statistics\/}~{\em 7\/}(1), 70--76.

\bibitem[\protect\citeauthoryear{Mahoney and Drineas}{Mahoney and
  Drineas}{2009}]{mahoney2009cur}
Mahoney, M.~W. and P.~Drineas (2009).
\newblock {CUR} matrix decompositions for improved data analysis.
\newblock {\em Proceedings of the National Academy of Sciences\/}~{\em
  106\/}(3), 697--702.

\bibitem[\protect\citeauthoryear{Mahoney, Maggioni, and Drineas}{Mahoney
  et~al.}{2008}]{mahoney2008tensor}
Mahoney, M.~W., M.~Maggioni, and P.~Drineas (2008).
\newblock Tensor-{CUR} decompositions for tensor-based data.
\newblock {\em SIAM Journal on Matrix Analysis and Applications\/}~{\em
  30\/}(3), 957--987.

\bibitem[\protect\citeauthoryear{Musco and Musco}{Musco and
  Musco}{2015}]{musco2015randomized}
Musco, C. and C.~Musco (2015).
\newblock Randomized block {Krylov} methods for stronger and faster approximate
  singular value decomposition.
\newblock In {\em Advances in Neural Information Processing Systems}, pp.\
  1396--1404.

\bibitem[\protect\citeauthoryear{Mutlu, Ural, and G{\"u}nd{\"u}z}{Mutlu
  et~al.}{2012}]{mutlu2012differential}
Mutlu, P., A.~U. Ural, and U.~G{\"u}nd{\"u}z (2012).
\newblock Differential gene expression analysis related to extracellular matrix
  components in drug-resistant {RPMI}-8226 cell line.
\newblock {\em Biomedicine \& Pharmacotherapy\/}~{\em 66\/}(3), 228--231.

\bibitem[\protect\citeauthoryear{Pandolfi}{Pandolfi}{2004}]{pandolfi2004aberrant}
Pandolfi, P.~P. (2004).
\newblock Aberrant {mRNA} translation in cancer pathogenesis: an old concept
  revisited comes finally of age.
\newblock {\em Oncogene\/}~{\em 23\/}(18), 3134--3137.

\bibitem[\protect\citeauthoryear{Ravikumar, Lafferty, Liu, and
  Wasserman}{Ravikumar et~al.}{2009}]{ravikumar2009sparse}
Ravikumar, P., J.~Lafferty, H.~Liu, and L.~Wasserman (2009).
\newblock Sparse additive models.
\newblock {\em Journal of the Royal Statistical Society: Series B (Statistical
  Methodology)\/}~{\em 71\/}(5), 1009--1030.

\bibitem[\protect\citeauthoryear{Shamir}{Shamir}{2016}]{shamir2016fast}
Shamir, O. (2016).
\newblock Fast stochastic algorithms for {SVD} and {PCA}: Convergence
  properties and convexity.
\newblock In {\em International Conference on Machine Learning}, pp.\
  248--256.

\bibitem[\protect\citeauthoryear{Shen, Shen, and Marron}{Shen
  et~al.}{2014}]{shen2014general}
Shen, D., H.~Shen, and J.~Marron (2014).
\newblock A general framework for consistency of principal component analysis.
\newblock {\em Journal of Machine Learning Research\/}~{\em 17\/}(150), 1--34.

\bibitem[\protect\citeauthoryear{Shen, Shen, Zhu, and Marron}{Shen
  et~al.}{2016}]{shen2016statistics}
Shen, D., H.~Shen, H.~Zhu, and J.~Marron (2016).
\newblock The statistics and mathematics of high dimension low sample size
  asymptotics.
\newblock {\em Statistica Sinica\/}~{\em 26\/}(4), 1747.

\bibitem[\protect\citeauthoryear{Siddiqui and Borden}{Siddiqui and
  Borden}{2012}]{siddiqui2012mrna}
Siddiqui, N. and K.~L. Borden (2012).
\newblock {mRNA} export and cancer.
\newblock {\em Wiley Interdisciplinary Reviews: {RNA}\/}~{\em 3\/}(1), 13--25.

\bibitem[\protect\citeauthoryear{St{\aa}hl, Salm{\'e}n, Vickovic, Lundmark,
  Navarro, Magnusson, Giacomello, Asp, Westholm, Huss, Mollbrink, Linnarsson,
  Codeluppi, Borg, Pont{\'e}n, Costea, Sahl{\'e}n, Mulder, Bergmann, Lundeberg,
  and Fris{\'e}n}{St{\aa}hl et~al.}{2016}]{staahl2016visualization}
St{\aa}hl, P.~L., F.~Salm{\'e}n, S.~Vickovic, A.~Lundmark, J.~F. Navarro,
  J.~Magnusson, S.~Giacomello, M.~Asp, J.~O. Westholm, M.~Huss, A.~Mollbrink,
  S.~Linnarsson, S.~Codeluppi, {\AA}.~Borg, F.~Pont{\'e}n, P.~I. Costea,
  P.~Sahl{\'e}n, J.~Mulder, O.~Bergmann, J.~Lundeberg, and J.~Fris{\'e}n
  (2016).
\newblock Visualization and analysis of gene expression in tissue sections by
  spatial transcriptomics.
\newblock {\em Science\/}~{\em 353\/}(6294), 78--82.

\bibitem[\protect\citeauthoryear{Stewart}{Stewart}{1998}]{stewart1998four}
Stewart, G. (1998).
\newblock Four algorithms for the efficient computation of truncated pivoted
  {QR} approximation to a sparse matrix. {CS} report.
\newblock Technical report, TR-98-12, University of Maryland.

\bibitem[\protect\citeauthoryear{Tibshirani}{Tibshirani}{1996}]{tibshirani1996regression}
Tibshirani, R. (1996).
\newblock Regression shrinkage and selection via the lasso.
\newblock {\em Journal of the Royal Statistical Society: Series B
  (Methodological)\/}~{\em 58\/}(1), 267--288.

\bibitem[\protect\citeauthoryear{Wang}{Wang}{2009}]{wang2009forward}
Wang, H. (2009).
\newblock Forward regression for ultra-high dimensional variable screening.
\newblock {\em Journal of the American Statistical Association\/}~{\em
  104\/}(488), 1512--1524.

\bibitem[\protect\citeauthoryear{Wu and Qu}{Wu and Qu}{2015}]{wu2015cancer}
Wu, L. and X.~Qu (2015).
\newblock Cancer biomarker detection: recent achievements and challenges.
\newblock {\em Chemical Society Reviews\/}~{\em 44\/}(10), 2963--2997.

\bibitem[\protect\citeauthoryear{Yuan and Lin}{Yuan and
  Lin}{2007}]{yuan2007non}
Yuan, M. and Y.~Lin (2007).
\newblock On the non-negative garrotte estimator.
\newblock {\em Journal of the Royal Statistical Society: Series B (Statistical
  Methodology)\/}~{\em 69\/}(2), 143--161.

\bibitem[\protect\citeauthoryear{Zeng and Zhu}{Zeng and
  Zhu}{2010}]{zeng2010integral}
Zeng, P. and Y.~Zhu (2010).
\newblock An integral transform method for estimating the central mean and
  central subspaces.
\newblock {\em Journal of Multivariate Analysis\/}~{\em 101\/}(1), 271--290.

\bibitem[\protect\citeauthoryear{Zhang}{Zhang}{2011}]{zhang2011matrix}
Zhang, F. (2011).
\newblock {\em Matrix theory: basic results and techniques}.
\newblock Springer Science \& Business Media.

\bibitem[\protect\citeauthoryear{Zhong, Zhang, Zhu, and Liu}{Zhong
  et~al.}{2012}]{zhong2012correlation}
Zhong, W., T.~Zhang, Y.~Zhu, and J.~S. Liu (2012).
\newblock Correlation pursuit: forward stepwise variable selection for index
  models.
\newblock {\em Journal of the Royal Statistical Society: Series B (Statistical
  Methodology)\/}~{\em 74\/}(5), 849--870.

\bibitem[\protect\citeauthoryear{Zhou, Zhu, Xu, and Li}{Zhou
  et~al.}{2020}]{zhou2019model}
Zhou, T., L.~Zhu, C.~Xu, and R.~Li (2020).
\newblock Model-free forward screening via cumulative divergence.
\newblock {\em Journal of the American Statistical Association\/}~{\em
  115\/}(531), 1393--1405.

\bibitem[\protect\citeauthoryear{Zhu, Miao, and Peng}{Zhu
  et~al.}{2006}]{zhu2006sliced}
Zhu, L., B.~Miao, and H.~Peng (2006).
\newblock On sliced inverse regression with high-dimensional covariates.
\newblock {\em Journal of the American Statistical Association\/}~{\em
  101\/}(474), 630--643.

\bibitem[\protect\citeauthoryear{Zhu, Li, Li, and Zhu}{Zhu
  et~al.}{2011}]{zhu2011model}
Zhu, L.-P., L.~Li, R.~Li, and L.-X. Zhu (2011).
\newblock Model-free feature screening for ultrahigh-dimensional data.
\newblock {\em Journal of the American Statistical Association\/}~{\em
  106\/}(496), 1464--1475.

\bibitem[\protect\citeauthoryear{Zou and Hastie}{Zou and
  Hastie}{2005}]{zou2005regularization}
Zou, H. and T.~Hastie (2005).
\newblock Regularization and variable selection via the elastic net.
\newblock {\em Journal of the Royal Statistical Society: Series B (Statistical
  Methodology)\/}~{\em 67\/}(2), 301--320.

\bibitem[\protect\citeauthoryear{Zou and Li}{Zou and Li}{2008}]{zou2008one}
Zou, H. and R.~Li (2008).
\newblock One-step sparse estimates in nonconcave penalized likelihood models.
\newblock {\em Annals of Statistics\/}~{\em 36\/}(4), 1509.

\end{thebibliography}

\end{document}